\documentclass[aps,prb,showpacs,twocolumn,floats,epsfig,pdflatex,longbibliography]{revtex4-2}
\usepackage{amssymb}
\usepackage{amsmath}
\usepackage{amsfonts}
\usepackage{graphicx}
\usepackage{amssymb}
\usepackage{float}
\usepackage{hyperref}
\usepackage{bbm} 
\usepackage{subcaption}
\usepackage{physics}

\usepackage{ragged2e}

\begin{document}

\title{Weak-coupling altermagnetism and chiral magnetic excitations in a checkerboard lattice}
\author{Abhigya Rangari}
\author{Manna Paul}
\author{Sayandip Ghosh}
\email{sayandipghosh@phy.vnit.ac.in}

\affiliation{Department of Physics, Visvesvaraya National Institute of Technology Nagpur, Nagpur 440010, India}

\date{\today}

\begin{abstract}
Altermagnets, characterized by spin-split electronic bands with compensated magnetic moments, have emerged as a new class of magnetic materials garnering attention in recent years. Here, using a minimal one-band Hubbard model, we show that the checkerboard lattice serves as a natural platform for altermagnetism for electrons. The instability towards altermagnetic order is denoted by diverging altermagnetic susceptibility at weak-coupling. Carrying out mean-field treatment of the Hubbard repulsion, we show phase
transitions from the nonmagnetic to altermagnetic semimetal and then to altermagnetic insulating phase, allowing clear identification of spin-split states. We then examine magnetic excitations in the altermagnetic phases using a random-phase approximation treatment of the dynamical spin susceptibility. The altermagnetic order is found to be stable against spin-fluctuations with the excitation spectra showing well-defined magnon excitations, which decay into single-particle excitations with decreasing interaction strength. Remarkably, the magnetic excitations exhibit strong dependence on both chirality and direction, showing an alternating chirality splitting, similar to the alternating spin splitting of the electronic bands, which serves as a salient feature of altermagnetism. 
\end{abstract}

\maketitle

\section{Introduction}
\label{Sec:intro}
Altermagnets have emerged as a new subclass of magnetic materials that is distinct from conventional ferromagnets and antiferromagnets ~\cite{smejkal2022emerging, smejkal2022beyond, xiao2024spin, jiang2024enumeration, chen2024enumeration, gao2025ai, shuai2026extremely, tamang2024newly}. Altermagnets exhibit zero net magnetization, similar to antiferromagnets, while simultaneously displaying momentum-dependent spin splitting in the electronic band structure, reminiscent of ferromagnets. Moreover, the opposite magnetic sublattices in an altermagnet are not related by translation or inversion symmetry, but instead are related by crystal rotations. These properties make altermagnets fascinating both on fundamental and applied aspects. In addition to the unique band structure, altermagnets also have a non-trivial magnon spectrum with the dispersion of magnons of different chiralities split in an anisotropic way, governed by symmetries. Also, their fast dynamics and symmetry driven transport make them ideal for memory devices, logic gates and sensors~\cite{beida2025chiral}. A growing number of material candidates have been identified to host altermagnetism, including RuO$_2$~\cite{fedchenko2024observation, guo2024direct}, MnF$_2$~\cite{yuan2020giant, bhowal2024ferroically}, MnTe~\cite{lee2024broken, osumi2024observation, krempasky2024altermagnetic}, CrSb~\cite{reimers2024direct,ding2024large, li2025topological}, FeSb$_2$~\cite{mazin2021prediction}, GdAlSi~\cite{nag2024GdAlSi} and others~\cite{guo2023spin}. The characteristic electronic structure of altermagnets gives rise to a variety of novel physical phenomena, such as the crystal Hall effect \cite{attias2024intrinsic,smejkal2020crystal,feng2022anomalous}, spin currents \cite{cui2023efficient}, crystal thermal transport \cite{zhou2024crystal}, magneto-optical Kerr effect \cite{solovyev1997magneto-optical}, spin–orbit torque \cite{vakili2025spin}, etc.

Earlier studies on altermagnetism were largely carried out by first-principle methods \cite{ahn2019antiferromagnetism,smejkal2020crystal,yuan2020giant,mazin2021prediction, gonzalez2021efficient, yuan2021prediction, egorov2021colossal, smejkal2022beyond}. More recently, electronic models in a variety of lattices such as square lattice~\cite{das2024realizing, li2025enhancement, ma2024altermagnetic, he2025altermagnetism}, square-octagon lattice~\cite{bose2024altermagnetism}, Lieb lattice and its variants~\cite{che2025engineering, wang2026spin, kaushal2025altermagnetism, biswas2026altermagnetic}, Shastry-Sutherland lattice~\cite{ferrari2024altermagnetism}, honeycomb lattice~\cite{sato2024altermagnetic}
demonstrated the existence of altermagnetic ground states. These studies using a range of analytical and numerical techniques exhibit the effect of crystal symmetries resulting in alternating spin polarization and spin-split band structures without net magnetization. In particular, Roig {\it et al.}  \cite{roig2024minimal} carried out a comprehensive symmetry analysis to develop realistic models for altermagnetism in nonsymmorphic materials. Moreover, altermagnets exhibit physics much richer than just alternating electronic bands, notably chiral splitting of magnon branches \cite{smejkal2023chiral}. Spin models have been employed to study this magnon chiral splitting~\cite{consoli2025altermagnetism,yershov2024fluctuation, wiedmann2025quantum} as well as to construct altermagnetic ordered state~\cite{zhu2025design,vijayvargia2025altermagnets}. Despite these advances, comparatively little attention paid to the nature of magnetic excitations in the itinerant systems. 

Recently, Maier {\it et al.} \cite{maier2023weak} explored the altermagnetic ordered state in a square lattice Hubbard model with orthorhombic anisotropy and studied dynamic spin susceptibility. They found that the magnetic excitation spectrum depends on its chirality, for momenta along certain directions in the Brillouin zone. However, the origin of the altermagnetic order was not addressed. In this work, we study the origin of altermagnetism and its associated magnetic excitations within a weak-coupling framework using a Hubbard model on a checkerboard lattice.

Spin models in the checkerboard lattice, also known as the planar-pyrochlore lattice, has long been extensively analyzed as it provides a quintessential platform to study frustrated magnetism ~\cite{lieb1999ground,palmer2000order,palmer2001quantum,starykh2002spinons, canals2002from, fouet2003planar, berg2003singlet, Tchernyshyov2003bond, bernier2004planar, moessner2004planar}. Lately, it has received considerable attention~\cite{consoli2025altermagnetism,yershov2024fluctuation, daghofer2026altermagnetic} as the magnon spectra in this lattice exhibit momentum dependent chiral splitting characteristic of altermagnets. Moreover, when the local spins interact with itinerant electrons via Kondo-type exchange couplings, the resulting electronic bands show altermagnetic splitting. 

In this article, we show that itinerant electrons on checkerboard lattices possess instability towards altermagnetic ordered state. We will first consider a minimal one-band Hubbard model on a checkerboard lattice (in Sec.~\ref{Sec:Model}) and investigate the momentum-dependent (static) spin susceptibility and altermagnetic susceptibility (in Sec.~\ref{Sec:sus_NSM}). The latter is defined in terms of sublattice-resolved spin operators and its uniform part features a peak indicating proclivity towards weak-coupling altermagnetism. In Sec.~\ref{Sec:Hartree-Fock}, we employ self-consistent Hartree-Fock theory to capture the momentum-dependent spin-splitting of the electronic bands. The system undergoes phase transitions to a semimetallic and then to an insulating phase as a function of the interaction strength $U$. We then examine magnetic excitations in the altermagnetic phase using the random phase approximation in Sec.~\ref{Sec:Magnetic_excitations} and identify chirality-dependent spin-wave excitations. In particular, the altermagnetic order is found to be robust against spin-fluctuations in the both semimetallic and insulating phases. Finally, we summarize our results in Sec.~\ref{Sec:Summary} and present some additional details in the Appendices.

\section{Model}
\label{Sec:Model}
\begin{figure}[t]
    \includegraphics[width=\linewidth]{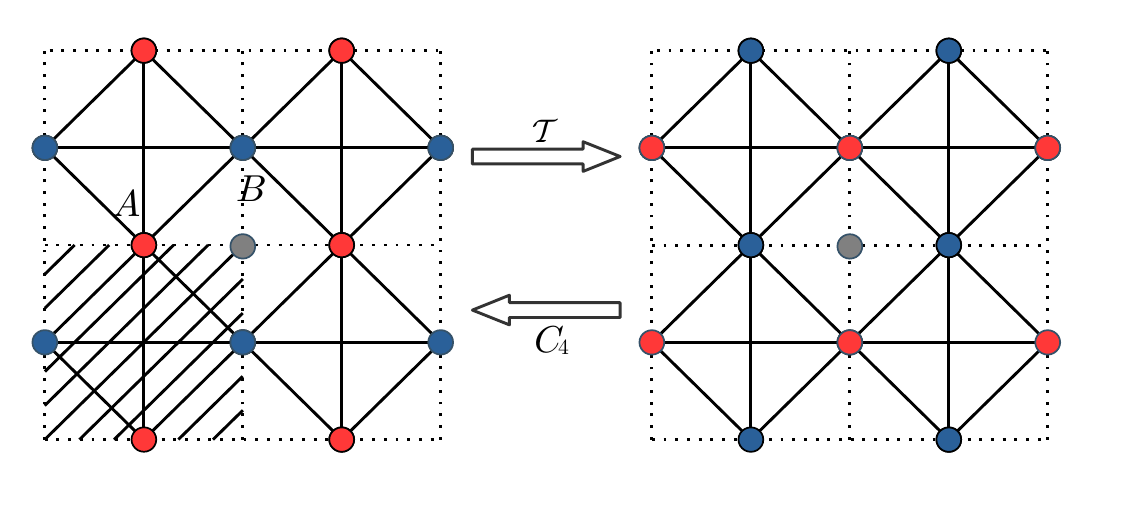}
  \caption{\justifying The checkerboard lattice with sublattices $A$ and $B$ marked by red and blue circles, respectively (left panel). The shaded area denotes the unit cell. The operation of time-reversal and $C_4$ real-space rotation around the plaquette center (gray circle) converts it to the dual lattice in the right panel. The lattice constant is taken to be unity.}
  \label{Fig:lattice}
\end{figure}

We start with two-dimensional (2D) Hubbard model on a checkerboard lattice shown in Fig.~\ref{Fig:lattice}. Its Hamiltonian reads
\begin{align}
{\cal H} = \sum_{i,j,\sigma} (t_{ij} - \mu \delta_{ij}) c_{i \sigma}^\dagger c_{j \sigma} + U \sum_{i} n_{i,\uparrow} n_{i,\downarrow},
\label{Eq:Hamil1}
\end{align}
where Latin indices $i$ and $j$ denote the sites of a 2D checkerboard lattice (which is not a Bravais lattice). Here, $c^\dagger_{i,\sigma} (c_{i,\sigma})$ creates (annihilates) an electron with spin $\sigma(\uparrow / \downarrow)$ on site $i$ with the corresponding number operator $n_{i,\sigma}\equiv c^\dagger_{i,\sigma} c_{i,\sigma}$ , $t_{ij}$ represents the hopping amplitude between sites $i$ and $j$. $U$ is the onsite Coulomb repulsion and $\mu$ the chemical potential. In this work, we consider hopping upto third-neighbor sites. Unless specified otherwise, the parameters are fixed at $t_1 = 1.0$ (defining the unit of energy), $t_2 = 0.4$, and $t_3 = 0.1$. The chemical potential $\mu$ is tuned to achieve half-filling.

In the N{\'e}el ordered state discussed later, the sublattices are related by time-reversal (${\cal T}$) as well as $C_4$ rotation in real-space around the plaquette center. Hence, the Hamiltonian symmetry-broken state is invariant under the combined operation of time-reversal followed by $C_4$ rotation.

Fourier transforming the non-interacting (first) term of Eq.~(\ref{Eq:Hamil1}) in reciprocal space, we obtain the tight-binding Hamiltonian
\begin{align}
{\cal H}_0 = \sum_{{\bf k},\sigma} \Psi^\dagger_{{\bf k},\sigma} h_0 ({\bf k}) \Psi_{{\bf k},\sigma},
\label{Eq:Hamil2}
\end{align}
with $ \Psi_{{\bf k},\sigma} = ( c_{A,{\bf k},\sigma} \ \ c_{B,{\bf k},\sigma})^T$ annihilates the two-component spinor state at crystal momentum ${\bf k}$ and 
\begin{align}
h_0 ({\bf k}) =  \left( \frac{\epsilon_{\bf k}^{2x} + \epsilon_{\bf k}^{2y}}{2} + \epsilon_{\bf k}^{3} - \mu \right)  \tau_0 + \epsilon_{\bf k}^1 \tau_x +  \frac{\epsilon_{\bf k}^{2x} - \epsilon_{\bf k}^{2y}}{2} \tau_z.  
\label{Eq:Hamil3}
\end{align}
Here, we have defined
\begin{align}
& \epsilon_{\bf k}^1 = 4 t_1 \cos \frac{k_x}{2} \cos \frac{k_y}{2}, \notag \\
& \epsilon_{\bf k}^{2x(2y)} = 2 t_2 \cos k_{x(y)}, \notag \\
& \epsilon_{\bf k}^3 = 4 t_3 \cos k_x \cos k_y,
\end{align} 
and $\tau_{0,x,y,z}$ are Pauli matrices in sublattice space. The two electronic energy bands are given as
\begin{align}
E_{\pm} ({\bf k}) =  {\cal E}_{0,{\bf k}}  \pm \sqrt{{\cal E}_{x,{\bf k}}^2 + {\cal E}_{z,{\bf k}}^2} - \mu, 
\label{Eq:Energy1}
\end{align}
where we have defined ${\cal E}_{0,{\bf k}}=\dfrac{\epsilon_{\bf k}^{2x} + \epsilon_{\bf k}^{2y}}{2} + \epsilon_{\bf k}^{3}$, ${\cal E}_{x,{\bf k}}=\epsilon_{\bf k}^1$, and ${\cal E}_{z,{\bf k}} = \dfrac{\epsilon_{\bf k}^{2x} - \epsilon_{\bf k}^{2y}}{2}$. The band structure along the Brillouin zone (BZ) symmetry direction are shown in Fig.~\ref{Fig:Band_AM}(a). The corresponding eigenfunctions (periodic parts of Bloch states) are given by
\begin{align}
|\phi_+ ({\bf k})\rangle &= \left( \begin{array}{c}
\cos \frac{\theta_{\bf k}}{2} \\ \sin \frac{\theta_{\bf k}}{2}
\end{array}\right), \qquad
|\phi_- ({\bf k})\rangle &= \left( \begin{array}{c}
\sin \frac{\theta_{\bf k}}{2} \\ - \cos \frac{\theta_{\bf k}}{2}
\end{array}\right), 
\end{align}
with $\theta_{\bf k} =  \arccos \left[ \frac{{\cal E}_{z,{\bf k}}}{\sqrt{{\cal E}^2_{x,{\bf k}} + {\cal E}^2_{z,{\bf k}}}}\right]$. 

\section{Susceptibilities and Altermagnetic Instability}
\label{Sec:sus_NSM}
To examine the proclivity of the system towards magnetic order, we look at magnetic susceptibilities in the nonmagnetic state. The (usual) spin susceptibility reads
\begin{align}
\chi^{\rm S} ({\bf q},i \omega_n) = \frac{1}{N}\int_0^{\beta} d\tau e^{i \omega_n \tau} \langle T_{\tau} {\bf S}({\bf q},\tau){\bf S}(-{\bf q},0)\rangle,
\end{align}
where ${\bf S}({\bf q}) = \sum_{{\bf k},\sigma} \Psi_{{\bf k} + {\bf q},\sigma}^\dagger \tau_0 \Psi_{{\bf k},\sigma}$ is the Fourier-transformed spin operator related to
the spin operator at site $i$ by
\begin{align}
{\bf S}_i = \frac{1}{\sqrt{N}}\sum_{\bf q} {\bf S}({\bf q}) e^{i {\bf q}\cdot {\bf r}_i}.
\end{align}
On the other hand, the spin susceptibility in the altermagnetic channel (which we call {\it altermagnetic susceptibility} throughout this work) is given by~\cite{roig2024minimal}
\begin{align}
\chi^{\rm AM} ({\bf q},i \omega_n) = \frac{1}{N}\int_0^{\beta} d\tau e^{i \omega_n \tau} \langle T_{\tau} {\tilde {\bf S}}({\bf q},\tau){\tilde {\bf S}}(-{\bf q},0)\rangle,
\end{align}
where ${\tilde {\bf S}}({\bf q}) = \sum_{{\bf k},\sigma} \Psi_{{\bf k} + {\bf q},\sigma}^\dagger \tau_z \Psi_{{\bf k},\sigma}$. Here, we have followed standard notation with $T_{\tau}$ being the time-ordering directive in imaginary time $\tau$, $\beta = (k_B T)^{-1}$ the inverse temperature, and $i \omega_n$ are bosonic Matsubara frequencies. These susceptibilities are obtained from the $2\times 2 \times 2\times 2$ generalized susceptibility tensor in the sublattice space whose elements are given as
\begin{align}
\chi_{\mu \mu' \nu \nu'} &({\bf q},i \omega_n) =  \frac{1}{N} \sum_{{\bf k},\sigma} \int_0^\beta d\tau e^{i \omega_n \tau} \\ \notag &\times \langle T_{\tau} c^\dagger_{\mu,{\bf k}+{\bf q},\sigma}({\tau})c_{\mu',{\bf k},\sigma}({\tau}) c^\dagger_{\nu,{\bf k},\sigma}(0) c_{\nu',{\bf k}+{\bf q},\sigma}(0)\rangle.
\end{align}
From now on, we will suppress the wave vector and frequency arguments for brevity unless necessary. In the random phase approximation (RPA), elements of the sublattice-dependent susceptibility read
\begin{align}
\chi^{\rm RPA}_{\mu \mu' \nu \nu'} &= \left[ \chi^{(0)}  \left( \mathbbm{1}- U \chi^{(0)} \right)^{-1} \right]_{\mu \mu' \nu \nu'},
\end{align}
with $\chi^{(0)}$ being the bare (non-interacting) susceptibility whose elements are described by
\begin{align}
\chi^{(0)}_{\mu \mu' \nu \nu'}& = - \frac{1}{N} \sum_{{\bf k},s,s'} \phi^*_{\mu,s'}({\bf k}+{\bf q})\phi_{\nu',s'}({\bf k}+{\bf q}) \\ \notag
&\times \phi_{\mu',s}({\bf k})\phi^*_{\nu,s}({\bf k})  \frac{n^F_s({\bf k})-n^F_{s'} ({\bf k}+{\bf q})}{i \omega_n + E_s({\bf k}) - E_{s'}({\bf k}+{\bf q}) }.
\end{align}
Here, $s,s'$ refer to the band indices with energies $E_{s/s'}$, $\phi_{\mu,s}$ is component of eigenfunction of the tight-binding Hamiltonian ${\cal H}_0$ [in Eq.~(\ref{Eq:Hamil2})] for sublattice $\mu$.  $n^F_s$ is the Fermi-Dirac distribution function for band $s$.

Carrying out analytic continuation $i \omega_n \to \omega + i \delta$, with $\omega$ being real frequency and $\delta$ a positive infinitesimal, the spin and altermagnetic susceptibilities are obtained as
\begin{align}
\chi^{\rm {S, RPA}}({\bf q},\omega) &= \sum_{\mu,\nu}  \chi^{\rm RPA}_{\mu \mu \nu \nu} ({\bf q},\omega + i \delta), \notag \\ 
\chi^{\rm {AM, RPA}}({\bf q},\omega) &= \sum_{\mu,\nu} (-1)^\mu (-1)^\nu \chi^{\rm RPA}_{\mu \mu \nu \nu} ({\bf q},\omega + i \delta).
\end{align}
Divergence of these susceptibilities in the static ($\omega \to 0$) limit indicates proclivity of the system towards magnetically ordered phase. If $\chi^{\mathrm{AM,RPA}}$ diverges before $\chi^{\mathrm{S,RPA}}$, the altermagnetic state is favored over ferromagnetic or spin-density-wave (SDW) ordering.

\begin{figure}[t]
  \begin{subfigure}{0.4\textwidth}
    \raggedright
    \caption{}
    \includegraphics[width=\linewidth]{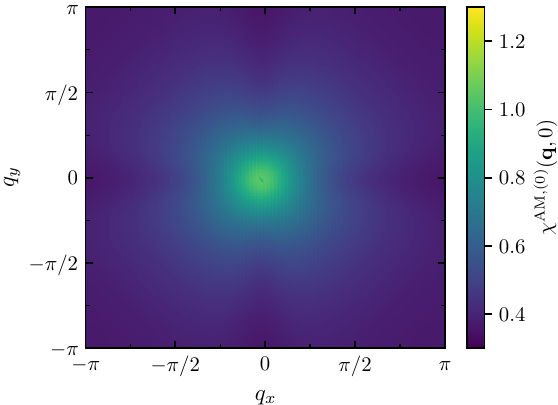}
  \end{subfigure}
  \vspace{0.0em}
  \begin{subfigure}{0.4\textwidth}
    \raggedright
    \caption{}
    \includegraphics[width=\linewidth]{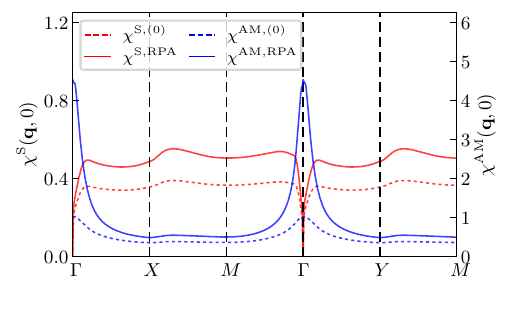}
  \end{subfigure}
\captionsetup{justification=raggedright,singlelinecheck=false}
\caption{\justifying
(a) The static bare altermagnetic susceptibility $\chi^{\rm AM,(0)}({\bf q})$ vs ${\bf q}$. 
(b) The static bare and RPA susceptibilities in the spin and altermagnetic channels for 
$U = 1.5$ and $T = 0.01$ along BZ symmetry directions 
$(0,0)\rightarrow (\pi,0)\rightarrow (\pi,\pi)\rightarrow (0,0)\rightarrow (0,\pi)\rightarrow (\pi,\pi)$. 
Note the differing vertical scales.
}
  \label{Fig:sus_spinless}
\end{figure}

Figure~\ref{Fig:sus_spinless}(a) shows the variation of bare altermagnetic susceptibility $\chi^{\rm AM,(0)}({\bf q},0)$ in the BZ exhibiting hotspots around the BZ center. Fig.~\ref{Fig:sus_spinless}(b) shows the (static) bare and RPA susceptibilities in spin and altermagnetic channels along paths connecting the high
symmetry points of the Brillouin zone (BZ): $\Gamma (0,0) \rightarrow X (\pi,0) \rightarrow M (\pi,\pi)\rightarrow \Gamma (0,0) \rightarrow Y (0,\pi)\rightarrow M(\pi,\pi) $. The static spin susceptibility shows very little variation in the  BZ indicating absence of instability towards ferromagnetic or SDW order, whereas the altermagnetic susceptibility shows divergence for ${\bf q}\to 0$ which suggests a transition to an altermagnetic phase at some critical interaction strength. Moreover, the band structure [in Fig.~\ref{Fig:Band_AM}(a)] at half-filling consists of point-like Fermi surface at BZ $M$ point ruling out nesting as underlying mechanism behind the altermagnetic instability.  All susceptibility calculations in this work are performed for $\delta = 0.001$ and $1000 \times 1000$ $\mathbf{k}$-point mesh. Smaller values of $\delta$ and ﬁner $\mathbf{k}$-point meshes do not produce qualitative or significant quantitative changes in our results.

\section{Hartree-Fock approximation}
\label{Sec:Hartree-Fock}
To further corroborate altermagnetic instability of electrons in our model, we now carry out a Hartree-Fock treatment of the Hubbard interaction term in Eq.~(\ref{Eq:Hamil1}). Introducing mean-field decoupling
\begin{align}
U n_{i,\uparrow} n_{i,\downarrow} \to U \left( \langle n_{i,\uparrow}\rangle n_{i,\downarrow} + n_{i,\uparrow} \langle n_{i,\downarrow}\rangle - \langle n_{i,\uparrow} \rangle \langle n_{i,\downarrow}\rangle \right),
\end{align}
and altermagnetic order parameter 
\begin{align}
m = \frac{1}{2N}\sum_{l} \left( n_{l,A,\uparrow} - n_{l,A,\downarrow} - n_{l,B,\uparrow} + n_{l,B,\downarrow} \right),
\end{align}
we consider a self-consistent solution in which the electron density is uniform while the spin polarization alternates between the sublattices. At filling $n$, the occupation number reads
\begin{align}
\langle n_{l,\mu,\sigma} \rangle = \frac{1}{2}\left( n + (-1)^{\mu+\sigma} m \right).
\end{align}
Here, $l$ is the unit cell index, $\mu$ the sublattice, and $\sigma$ the spin. For the alternating sign of the order parameter $(-1)^{\mu + \sigma}$, we indicate $\mu$ and $\sigma$ with $0$ for $A$ and $\uparrow$ and with $1$ for $B$ and $\downarrow$, respectively. In other words, a nonzero order parameter $m$ refer to staggered magnetization of Néel ordering, which, combined with the lattice symmetries gives rise to the altermagnetic state. Unlike conventional antiferromagnetic order in Bravais lattice, where the magnetic ordering enlarges the unit cell through a finite ordering wave vector, the altermagnetic order here does not enlarge the unit cell. This is because the opposite spin sublattices are connected by a rotational symmetry rather than translation or inversion~\cite{spaldin2026why}.

Carrying out the mean-field approximation, the Fourier transformed Hamiltonian is given as
\begin{align}
{\cal H}_{\rm HF} = \sum_{{\bf k}} \Psi^\dagger_{{\bf k}} \left( h_0 ({\bf k}) \otimes \sigma_0 - \Delta \tau_z \otimes \sigma_z \right) \Psi_{{\bf k}},
\label{Eq:Hamil_HF}
\end{align}
with $ \Psi_{{\bf k}} = ( c_{A,{\bf k},\uparrow} \  c_{B,{\bf k},\uparrow} \ c_{A,{\bf k},\downarrow}  \ c_{B,{\bf k},\downarrow})^T$, $\sigma_{0,x,y,z}$ are Pauli matrices in spin space, and $\Delta = m U/2$ is the staggered exchange field. Here, we have absorbed the constant term $n U /2$ into the chemical potential. Evidently the Hamiltonian is block diagonal in the spin space, and consequently the bands are spin polarized as discussed later.

To self-consistently determine the order parameter $m$, we evaluate the Hamiltonian matrix in Eq.~(\ref{Eq:Hamil_HF}) over $1000 \times 1000$ ${\bf k}$-point mesh. The expectation value of the local density operator $\langle n_{\mu,\sigma} \rangle$ is then calculated using the Hamiltonian eigenstates and eigenvalues in the symmetry-broken phase at temperature $T=0.01$. Iterations by updating the exchange fields and the chemical potential while fixing the total number of electrons are performed until self-consistency is achieved.

\begin{figure}
\includegraphics[scale=0.8]{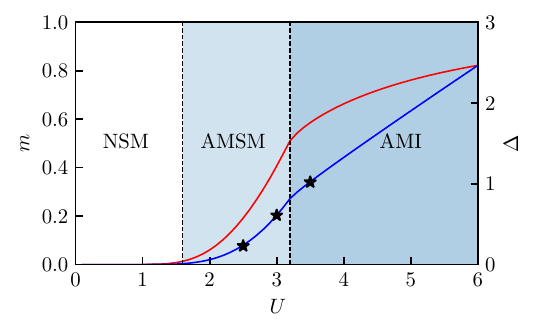}
\caption{\justifying The order parameter (red) and exchange field (blue) at half-filling from the self-consistent Hartree-Fock calculation, showing normal semimetal (NSM), altermagnetic semimetal (AMSM), and altermagnetic insulator (AMI) phases.}
\label{Fig:self_consistency}
\end{figure}

\begin{figure}
\includegraphics[scale=0.47]{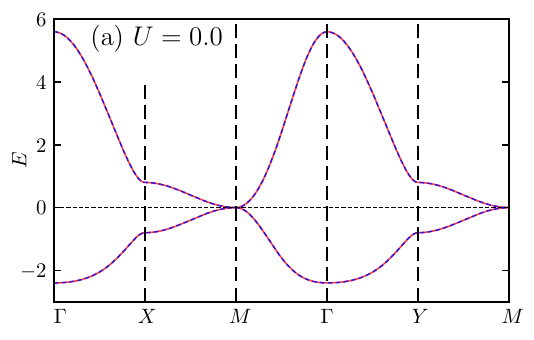}
\includegraphics[scale=0.47]{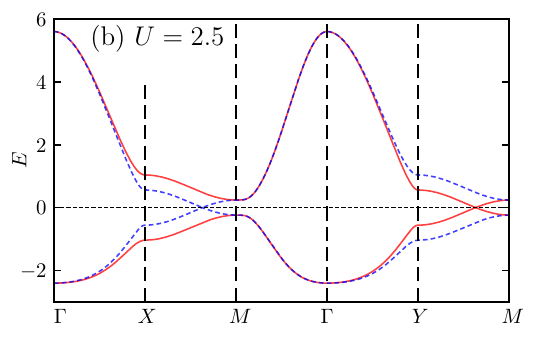}\\
\includegraphics[scale=0.47]{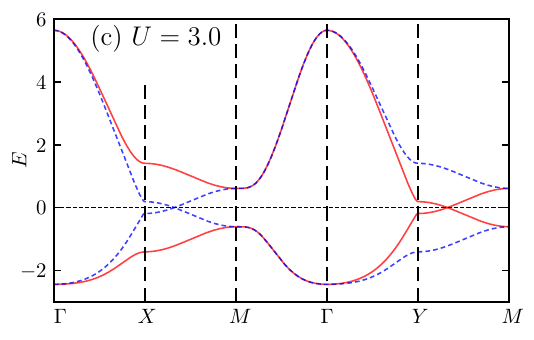}
\includegraphics[scale=0.47]{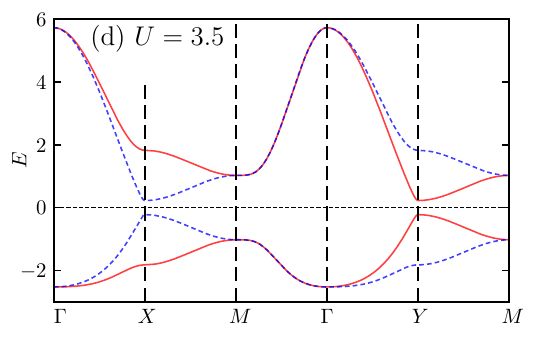}\\
\caption{\justifying Spin-resolved band structure from Eq.~(\ref{Eq:Hamil2}) for $U = 0$ (a), $U = 2.5$ (b), $U = 3.0$ (c), and $U = 3.5$ (d) with solid red and dashed blue lines denoting $\uparrow$ and $\downarrow$-spins, respectively.}
\label{Fig:Band_AM}
\end{figure}

Figure~\ref{Fig:self_consistency} shows the variation of order parameter and exchange field at half-filling $(n=1)$ with Coulomb repulsion $U$. The corresponding band structures at points marked by $\star$ is shown in Fig.~\ref{Fig:Band_AM}. For small $U$, we have the normal semimetallic (NSM) state with spin degenerate bands touching each other at $M$-point of the BZ. Above a threshold interaction strength $U \gtrsim 1.6$, a phase transition altermagnetic semimetallic (AMSM) state occurs with a sharp rise in $m$ and $\Delta$ values. The spin degeneracy of the band structure is lifted and the band touching points start moving towards $X$ and $Y$ points of the BZ. The band touching points are no longer spin-degenerate with $\uparrow$-spin ($\downarrow$-spin) bands being gapped for nonzero $k_x$ ($k_y$). However, spin degeneracy is not lifted along the $k_x = k_y$ line. The rapid increase of $m$ and $\Delta$ continues with increasing $U$ till a kink in the order parameter for $U \approx 3.2$. It indicates transition from AMSM to altermagnetic insulating (AMI) phase with direct band gaps opening at $X$ and $Y$ points. For larger $U$, the increase of $m$ and $\Delta$ slows down significantly. In both AMSM and AMI phases, the spin-degeneracy is lifted and splitting between up (red) and down (blue) depends on the crystal momentum ${\bf k}$. These features remain qualitatively robust against small variations in hopping parameters [see Appendix~\ref{ap:HF}], signaling a generic instability of itinerant electrons in checkerboard lattices toward altermagnetic order.

To understand the spin splitting, we take a closer look at the tight-binding Hamiltonian in Eq.~(\ref{Eq:Hamil3}). The $\frac{1}{2}\left( \epsilon_{\bf k}^{2x} - \epsilon_{\bf k}^{2y}\right) \tau_z$ term breaks ${\cal P}{\cal T}$ symmetry, with ${\cal P}$ being the inversion with respect to the plaquette center. In antiferromagnets, ${\cal P}{\cal T}$ symmetry preserves the spin degeneracy throughout the band structure. Consequently, its absence in our model lifts the degeneracy between the $\mathbf{k}\uparrow$ and $\mathbf{k}\downarrow$ states for general momenta. Spin splitting occurs everywhere except along the diagonal $k_x = k_y$ and at the BZ boundaries $\abs{k_x \pm k_y} = \pi$. Moreover, as sublattices are related by $C_4$ symmetry, the Hartree-Fock Hamiltonian in Eq.~(\ref{Eq:Hamil_HF}) is invariant under the combined operation of time-reversal followed by $C_4$ rotation in momentum space. In other words, the band structure possess a combined spin-flip-momentum-rotation: $k_x \leftrightarrow k_y; \uparrow \leftrightarrow \downarrow$ symmetry as evident in the Fig.~\ref{Fig:Band_AM}.
Accordingly, we obtain $d$-wave altermagnetic state in our model (see App.~\ref{ap:HF}). Furthermore, it should be noted that our Hartree-Fock Hamiltonian does not contain any spin-orbit-coupling indicating the non-relativistic nature of the altermagnetic band splitting.

\section{Magnetic excitations}
\label{Sec:Magnetic_excitations}

Now, we study itinerant magnetic excitations in our model. To this end, we look at the  transverse spin susceptibilities 
\begin{align}
\chi^{+-} ({\bf q},i \omega_n) = \frac{1}{N}\int_0^{\beta} d\tau e^{i \omega_n \tau} \langle T_{\tau} {\bf S}^+({\bf q},\tau){\bf S}^-(-{\bf q},0)\rangle, \notag \\
\chi^{-+} ({\bf q},i \omega_n) = \frac{1}{N}\int_0^{\beta} d\tau e^{i \omega_n \tau} \langle T_{\tau} {\bf S}^-({\bf q},\tau){\bf S}^+(-{\bf q},0)\rangle,
\end{align}
with ${\bf S}^+({\bf q}) = \sum_{{\bf k}} \Psi_{{\bf k} + {\bf q},\uparrow}^\dagger \Psi_{{\bf k},\downarrow}$ and ${\bf S}^-({\bf q}) = \sum_{{\bf k}} \Psi_{{\bf k} + {\bf q},\downarrow}^\dagger \Psi_{{\bf k},\uparrow}$ referring to spin raising and lowering operators, respectively. This dynamical susceptibility accounts for both collective and single-particle magnetic excitations. Following method similar to Sec.~\ref{Sec:sus_NSM}, the transverse susceptibilities are obtained from
\begin{align}
\chi^{\kappa} = \sum_{\mu,\nu} \chi^{\kappa}_{\mu \mu \nu \nu}. 
\end{align}
Here, ${\kappa} = + 1$ ($-1$) refer to $+-$ ($-+$), and the components of transverse susceptibility tensor are given by
\begin{align}
\chi^{\kappa}_{\mu \mu' \nu \nu'} &=
\frac{1}{N} \sum_{{\bf k}} \int_0^\beta d\tau e^{i \omega_n \tau}
\nonumber\\
&\times \langle T_{\tau}
c^\dagger_{\mu,{\bf k}+{\bf q},\sigma}({\tau})
c_{\mu',{\bf k},\bar{\sigma}}({\tau})
c^\dagger_{\nu,{\bf k},\bar{\sigma}}(0)
c_{\nu',{\bf k}+{\bf q},\sigma}(0)\rangle
\label{Eq:sus_RPA_AM},
\end{align}
with $\sigma (\bar{\sigma})$ refer to $\uparrow$ ($\downarrow$) spin for $\kappa = + 1$ and vice-versa for $\kappa = -1$. Similar to Sec.~\ref{Sec:sus_NSM}, the transverse  interacting susceptibilities in the RPA reads
\begin{align}
\chi^{\kappa,{\rm RPA}}_{\mu \mu' \nu \nu'}  &= \left[ \chi^{\kappa,(0)} \left( \mathbbm{1}- U \chi^{\kappa,(0)}\right)^{-1}\right]_{\mu \mu' \nu \nu'},
\label{Eq:RPA_magnetic}
\end{align}
with $\chi^{\kappa,(0)}$ being the mean-field susceptibilities calculated from eigenstates of the mean-field Hamiltonian [in Eq.~(\ref{Eq:Hamil_HF})]. The dynamical transverse susceptibility is obtained from the usual analytic continuation
\begin{align}
\chi^{\kappa}({\bf q},\omega) = \sum_{\mu,\nu} \chi^{\kappa,{\rm RPA}}_{\mu \mu \nu \nu} ({\bf q},\omega + i \delta). 
\end{align} 
The imaginary part of this susceptibility is related the spin-wave spectral functions whose peaks are associated with the spin-wave energies. The spin-wave dispersion can also be calculated from poles of Eq.~(\ref{Eq:RPA_magnetic}). Positive spin-wave energies in the entire BZ indicate stability of the altermagnetic phase against spin fluctuations. 

In ferromagnets (FMs), magnetic excitation corresponds to an electronic transition from the majority-spin band ($\uparrow$) to the minority-spin band ($\downarrow$). The electron-hole pair involved in this process form a bound state owing to el-el interaction $U$. This bound state, with wavevector-dependent frequency $\omega_{\bf q}$, is the magnon associated with spin-wave excitations. Magnons in FMs reduce the total magnetization by one quantum of angular momentum $\hbar$, and manifest as well-defined peaks in $\text{Im}(\chi^{+-}$), whereas  single-particle spin-flip excitations appear in the Stoner continuum. On other hand, in antiferromagnets (AFMs) magnetic excitations correspond to either lowering the spin on the majority $\uparrow$-spin sublattice, or raising the spin on the majority $\downarrow$-spin sublattice. Due to the translational symmetry between the sublattices, the two excitation spectra are degenerate across the Brillouin zone. In other words, the spectra for $\chi^{+-}$ and $\chi^{+-}$ are identical for AFMs. However, as shown below, the $\chi^{+-}$ and $\chi^{+-}$ in altermagnetic state in our model exhibit different spectra along momentum directions where electronic spin splitting occurs. Here, we adopt the convention that $\text{Im} (\chi^{+-})$ [$\text{Im} (\chi^{-+}$)] represents the spectral density of positive- (negative-) chirality magnons associated with the positive- (negative-) chirality neutron-scattering channel.

\begin{figure}[htb]
\centering
\includegraphics[scale=0.6]{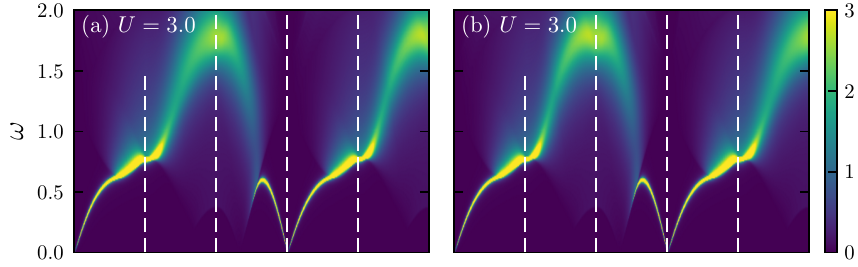}
\vspace{0mm}
\includegraphics[scale=0.6]{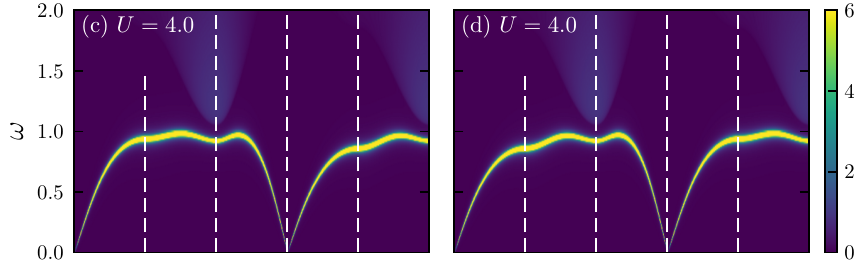}
\vspace{0mm}
\includegraphics[scale=0.6]{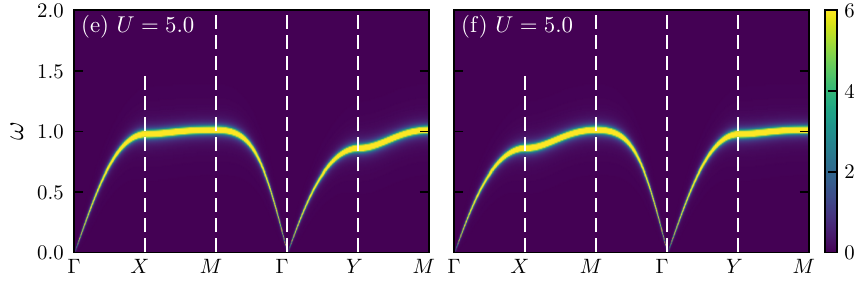}
\caption{\justifying Imaginary part of RPA transverse susceptibilities ($\chi^{+-}$ in the left panel and $\chi^{-+}$ in the right panel) for $U = 3.0$ (a)-(b), $U = 4.0$ (c)-(d), and $U = 5.0$ (e)-(f).}
\label{Fig:Sus_spinfull}
\end{figure}

The imaginary part of RPA transverse susceptibilities for different values of $U$ is shown in Fig.~\ref{Fig:Sus_spinfull} ($\chi^{+-}$ in the left panel and $\chi^{-+}$ in the right panel). For the semi-metallic case ($U=3.0$), we obtain propagating, gapless spin-wave excitations near the $\Gamma$-point. These Goldstone mode excitations stem from the absence of magnetic anisotropy in our model. At higher energies, the propagating spin waves obtain finite damping around the $X$ and $Y$ points, however, still with well-defined peaks. Around the $M$-point, the high-energy the spin-wave excitations decay into the Stoner continuum corresponding to the single-particle spin-flip excitations. In the insulating altermagnets ($U \gtrsim 3.2$), significant spectral weight of Stoner continuum is transferred to spin-wave excitations and we obtain undamped spin waves. It is important to note here that for intermediate coupling considered here, the sublattices are partially magnetized ($m < 1$), and one would naively expect the zone boundary spin-waves to get damped since their energies exceed the electronic band gap. However, the direct band gap around $X$ and $Y$ points occur for same-spin bands, and consequently we don't get any Stoner continuum corresponding to single-particle spin-flip excitations. The spin-degenerate band gap at $M$ point yields Stoner continuum into which the spin-wave excitations decay for $U=3.0$ when spin wave energies exceed the exchange gap $2\Delta$. For $U=4.0$, the Stoner continuum is visible above the spin wave spectrum. Positive spin wave energies in the entire BZ indicate stability of long-range order against spin-fluctuations in both semimetallic and insulating altermagnetic phase. 

Upon a closer look at the magnetic excitation spectra near $\Gamma$ point reveal that dispersion along $q_x$ from $\Gamma$ to $X$ is different from the dispersion along $q_y$ from $\Gamma$ to $Y$. This difference between the transverse susceptibilities along the BZ directions is opposite for $\chi^{+-}$ and $\chi^{-+}$. In other words, the spectral densities for positive chirality magnons along $q_x$ ($q_y$) is same as those for negative chirality magnons along $q_y$ ($q_x$).  Along the $\Gamma-M$ direction, where
electronic spin splitting is absent, the spectral densities are identical. This stems from the symmetry of the Hartree-Fock Hamiltonian under combined operation of time-reversal followed by $C_4$ rotation in momentum space. The combined spin-flip-momentum-rotation symmetry of the electronic states propagates to the magnon spectral densities as is apparent from Eq.~(\ref{Eq:sus_RPA_AM}). This demonstrates that magnetic excitations in altermagnets are intrinsically linked to their spin-dependent electronic structure.

\begin{figure}
\centering
\includegraphics[scale=0.8]{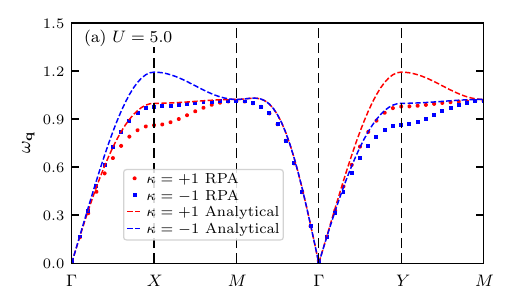}\\
\includegraphics[scale=0.8]{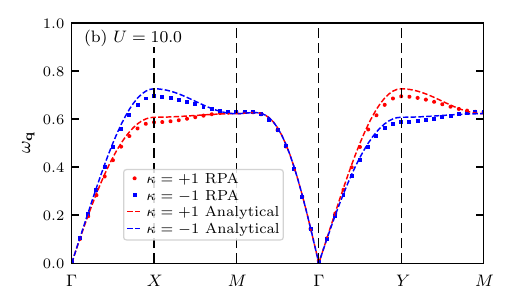}
\caption{\justifying Spin-wave dispersion in an insulating altermagnet in the intermediate ($U=5.0$) and strong ($U=10.0$) coupling regime. The dashed lines show the spin-wave dispersions obtained from the analytical expression [Eq.~(\ref{Eq:Analytical_StrongCoupling})] in the strong-coupling limit.}
\label{Fig:sw}
\end{figure}

To elucidate this point further, we plot the spin-wave dispersion for $U=5.0$ (intermediate coupling) and $U = 10.0$ (strong coupling) in Fig. \ref{Fig:sw}. Evidently, the two magnon branches exhibit have an alternating chirality splitting, akin to the alternating spin splitting of the electronic bands. The chiral splitting occurs for all momenta except along the $q_x = q_y$ line and at the BZ boundary $\abs{q_x \pm q_y} = \pi$. The chiral splitting ranges from $\sim 10 \%$ for intermediate coupling to $\sim 15 \%$ for strong-coupling. It is noteworthy that the long-wavelength spin-wave modes show linear ($\omega_{\bf q}\propto q$) behavior similar to AFM magnons. Furthermore, for typical hopping parameter value of $t_1 \sim 200$ meV for transition metals, the calculated spin-wave frequencies lies in THz range, akin to AFMs.

In the strong coupling ($U \gg t$) regime, suppression of charge fluctuations leads to electron localization and the system minimizes its energy through virtual hopping processes that favor antiparallel alignments of spins. Consequently, the Hubbard model reduces to the Heisenberg spin model with an effective antiferromagnetic exchange coupling $J= 4 t^2 / U$ \cite{Auerbach1994}. To this end, we carry out calculation of spin wave dispersion in the strong-coupling limit (see App. \ref{ap:strong_coupling}) in our model. The analytical expression of spin wave dispersion thus obtained read
\begin{align}
\frac{\omega_{{\bf q},\kappa}}{2 m J_1} &=   \sqrt{\left( 1 - \dfrac{J_2}{2J_1}\left( \gamma'_{q_x} + \gamma'_{q_y}\right) - \frac{J_3}{J_1}\gamma''_{\bf q}  \right)^2 - \gamma^2_{\bf q} } \notag \\ &-\kappa \dfrac{J_2}{2J_1}\left( \gamma'_{q_x} - \gamma'_{q_y}\right),
\label{Eq:Analytical_StrongCoupling}
\end{align}
with $J_i = {4t^2_i}/{U} \ (i=1,2,3)$, $\gamma_{\bf q} =  \cos \dfrac{q_x}{2} \cos \dfrac{q_y}{2}$, $\gamma'_{q_{x/y}} = \sin^2 \dfrac{q_{x/y}}{2}$, and $ \gamma''_{\bf q}=\left( 1 - \cos q_x \cos q_y \right)$. The calculated dispersion along BZ symmetry directions is shown in dashed line in Fig.~\ref{Fig:sw}. Predictably, the RPA treatment of Hubbard model is in good agreement with analytical expression in the strong-coupling case, whereas resulting in significant departures for intermediate coupling.  The same expression for spin wave dispersion can also be obtained for Holstein-Primakoff magnons in a Heisenberg spin model, including up-to third-neighbor exchange (see App.~\ref{sec:sw_heisenberg}).

The chiral splitting in our itinerant model arises from the anisotropic propagation of spin waves of opposite chirality. Projecting the spin-wave spectral function on the sublattices reveal that the $\kappa = + 1$ ($\kappa = - 1$) spin waves, particularly short-wavelength modes, propagate mostly on the $A$ ($B$) sublattice. The second neighbor hopping $t_2$ connecting $A$ ($B$) sublattice along the $x$ ($y$) direction in checkerboard lattice favors anisotropic propagation of the spin waves and produces such strong directionality which is rare for collective excitations. It is possible to exploit the chiral anisotropy to generate spin currents efficiently, which makes it promising for spintronics applications.

The chiral splitting of spin waves in altermagnets was predicted by first-principle calculations~\cite{smejkal2023chiral,zhang2025chiral,gonzalez2026chiral}, and demonstrated in inelastic neutron scattering studies~\cite{liu2024chiral,sun2025observation}. However, the explanation of the observed splitting requires unrealistic long-range exchange interactions in spin models~\cite{liu2024chiral,sun2025observation,hoyer2025altermagnetic}. In our electronic model, the splitting naturally arises due to anisotropy of the second-neighbor hopping on the checkerboard lattice. Similar chiral splitting has also been predicted in the square-lattice Hubbard model~\cite{costa2025giant} applicable to ultracold fermionic atoms in optical lattices~\cite{das2024realizing}. Incidentally, the chiral splitting in our model ranging $20 - 40$ meV~\footnote{We assume typical hopping parameter value of $t_1 \sim 200$ meV for transition metals} for intermediate-to-strong coupling is much larger than that predicted by first-principle calculations and linear-spin-wave theories~\cite{zhang2025chiral, liu2024chiral, hoyer2025altermagnetic}. Moreover, the splitting is not smeared by Stoner damping, thereby facilitating its experimental detection.     

\section{Summary and Outlook}
\label{Sec:Summary}
To summarize, we have shown that an altermagnetic ordered state with staggered magnetization naturally arises in a Hubbard model on checkerboard lattice owing to symmetry operation of time-reversal followed by $C_4$ rotation. The instability towards long-range order is characterized by diverging static susceptibility in the altermagnetic channel defined in terms of sublattice-resolved spin operators. Carrying out Hartree-Fock treatment of the Hubbard interaction $U$, we found phase transition from non-magnetic to altermagnetic state with momentum dependent spin-splitting of electronic bands which evolves from a semimetallic state at small $U$ to an insulating state at larger $U$. This provides us with theoretical insights of the underlying mechanism driving altermagnetism in itinerant models.

Using RPA, we have studied the dynamical spin susceptibility to analyze magnetic excitations in the altermagnetic state. The altermagnetic order is found to be stable against spin-fluctuations. At intermediate interaction strength, the magnetic excitation spectra exhibit well-defined magnon excitations which decay into single-particle excitations with decreasing interaction strength. Crucially, the excitation spectrum for opposite chirality spin-excitations exhibit alternating splitting across the Brillouin zone. This chiral splitting in our itinerant model emerges due to anisotropy of the second nearest neighbor hopping on the checkerboard lattice. The alternating sign of chiral splitting is a unique feature of altermagnets, having no equivalence from conventional ferromagnets and antiferromagnets, and therefore can be used as fingerprint of altermagnetism using inelastic neutron scattering studies. Our work thus provides a minimal framework to understand the origin of altermagnetism and chiral spin-wave excitations for electrons using checkerboard lattice as a prototypical platform.

The presence of chiral anisotropy together with the THz-range linear-dispersing magnons typical of antiferromagnets opens up new possibilities for spin caloritronics~\cite{cui2023efficient}. Moreover, they provide new pathways towards designing devices that are not only efficient and tunable but are also capable of functionalities beyond what conventional spintronics can achieve~\cite{smejkal2022anomalous, liu2025unconventional, jungwirth2025altermagneticspintronics}.

\section*{Acknowledgments}
The authors acknowledges financial support from the Anusandhan National Research Foundation (ANRF) under Grant No. ANRF/ECRG/2024/001412/PMS and Department of Science and Technology (DST) under DST-FIST Project No. SR/FST/PSI/2017/5(C).

%=========================Appendix======================================================

\begin{widetext}

\appendix
\section{Further Comments of the Hartree-Fock results}
\label{ap:HF}

\begin{figure}[ht]
    \centering
        % Left Subfigure
    \begin{minipage}[t]{0.48\textwidth}
        \subcaption{\label{Fig:mvsU}}
        \vspace{-1ex} 
        \includegraphics[width=\textwidth]{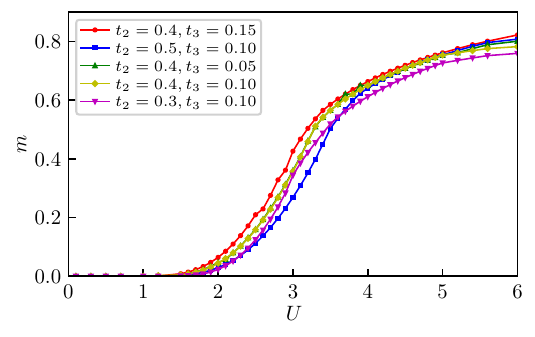}
    \end{minipage}
    \hfill 
    % Right Subfigure
    \begin{minipage}[t]{0.48\textwidth}
        \subcaption{\label{Fig:FS}} 
        \vspace{-3.5ex}
        \includegraphics[width=0.8\textwidth]{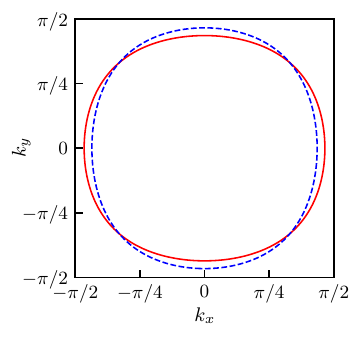}
    \end{minipage}

    \caption{\justifying (a) The order parameter from the self-consistent Hartree-Fock calculation for different values of $t_2$ and $t_3$. The value of $t_1$ is fixed at $1.0$. (b) Isoenergy contours at $\mu = 4.0$ and $U=3.5$. Solid red and dashed blue lines denote $\uparrow$ and $\downarrow$-spin states, respectively.}
    \label{Fig:HF_app}
\end{figure}

The variation of order parameter $m$ with interaction strength $U$ is shown in Fig.~\ref{Fig:mvsU} for different values of $t_2$ and $t_3$ keeping $t_1$ fixed at $1.0$. Evidently, the onset of staggered magnetization at $U \gtrsim 1.6$ and its subsequent increase with $U$ is insensitive to small changes in the hopping parameters, indicating stability of the altermagnetic order. The isoenergy contours for $\uparrow$-spin and $\downarrow$-spin states are shown in Fig.~\ref{Fig:FS}, demonstrating a $d$-wave symmetry.

\section{Spin-wave dispersion in the strong-coupling limit}
\label{ap:strong_coupling}
Is it well known that the Hubbard model in strong-coupling ($U \gg t$) limit is equivalent to Heisenberg spin model with effective spin-interaction $J = 4t^2/U$~\cite{Auerbach1994}. Here, we analytically calculate the magnon dispersion for our model in strong-coupling limit and show that the dispersion is identical to that obtained from a spin-model.

Rewriting the Hartree-Fock Hamiltonian ${\cal H}_{\rm HF} =  \sum_{{\bf k},\sigma} \Psi^\dagger_{{\bf k}} H_{\rm HF}({\bf k},\sigma) \Psi_{{\bf k}}$ with 
\begin{align}
H_{\rm HF}({\bf k},\sigma)= \begin{pmatrix}
{\cal E}_{0,{\bf k}} + \left( {\cal E}_{z,{\bf k}} - \sigma \Delta \right) & {\cal E}_{x,{\bf k}} \\
{\cal E}_{x,{\bf k}} & {\cal E}_{0,{\bf k}} - \left( {\cal E}_{z,{\bf k}} - \sigma \Delta \right)
\end{pmatrix}.
\label{Eq:ap_Hamil3}
\end{align}
Here, ${\cal E}_{0/x/z,{\bf k}}$ are defined in Sec.~\ref{Sec:Model}, and we have absorbed the chemical potential in ${\cal E}_{0,{\bf k}}$. The eigenvalues corresponding to the quasiparticle energies are
\begin{align}
E_{s} ({\bf k}, \sigma) = {\cal E}_{0,{\bf k}} + s \sqrt{{\cal E}^2_{x,{\bf k}} + \left( {\cal E}_{z,{\bf k}} - \sigma \Delta \right)^2},
\end{align}
with $s = \pm 1$ being the band-index. The corresponding eigenstates are
\begin{align}
|\phi_+ ({\bf k},\sigma)\rangle = \left( \begin{array}{c}
\cos \frac{\theta_{{\bf k},\sigma}}{2} \\ \sin \frac{\theta_{{\bf k},\sigma}}{2}
\end{array}\right), \;\;\;
|\phi_- ({\bf k},\sigma)\rangle = \left( \begin{array}{c}
\sin \frac{\theta_{{\bf k},\sigma}}{2} \\ - \cos \frac{\theta_{{\bf k},\sigma}}{2}
\end{array}\right), 
\end{align}
with $\theta_{{\bf k},\sigma} =  \arccos \left[ \dfrac{{\cal E}_{z,{\bf k}} - \sigma \Delta}{\sqrt{{\cal E}^2_{x,{\bf k}} + ({\cal E}_{z,{\bf k}} - \sigma \Delta)^2}}\right]$. 

As discussed in the main text, the magnon frequencies are obtained from the poles of RPA dynamical transverse susceptibilities in Eq.~(\ref{Eq:RPA_magnetic}). In RPA, only the intra-sublattice components $\chi^{\kappa,(0)}_{\mu \nu} $($\equiv \chi^{\kappa,(0)}_{\mu \mu \nu \nu}$) of mean-field transverse susceptibilities are relevant and they are given by
\begin{align}
\chi^{+-,(0)}_{\mu \nu}&({\bf q},i \omega_n) = - \frac{1}{N} \sum_{{\bf k},s,s'} \phi^*_{\mu,s'}({\bf k}+{\bf q},\uparrow) \phi_{\mu,s}({\bf k},\downarrow)\phi^*_{\nu,s}({\bf k},\downarrow) \phi_{\nu,s'}({\bf k}+{\bf q},\uparrow) \frac{n^F_s({\bf k})-n^F_{s'} ({\bf k}+{\bf q})}{i \omega_n + E_s({\bf k}) - E_{s'}({\bf k}+{\bf q}) },\notag \\
\chi^{-+,(0)}_{\mu \nu}&({\bf q},i \omega_n) = - \frac{1}{N} \sum_{{\bf k},s,s'} \phi^*_{\mu,s'}({\bf k}+{\bf q},\downarrow) \phi_{\mu,s}({\bf k},\uparrow)\phi^*_{\nu,s}({\bf k},\uparrow) \phi_{\nu,s'}({\bf k}+{\bf q},\downarrow) \frac{n^F_s({\bf k})-n^F_{s'} ({\bf k}+{\bf q})}{i \omega_n + E_s({\bf k}) - E_{s'}({\bf k}+{\bf q}) }.
\label{Eq:Sus_strong-coupling}
\end{align}
In the AMI phase, only interband ($s\neq s'$) processes contribute to the  mean-field susceptibilities. For arbitrary $U$, the ${\bf k}$-sum in the matrix elements in Eq.~(\ref{Eq:Sus_strong-coupling}) can be carried out numerically. In the strong-coupling limit, expanding these matrix elements in powers of $t_i/\Delta$ (${i=1,2,3}$) and $\omega/\Delta$, and systematically retaining terms only up to second order in $t_i/ \Delta$, and after performing the ${\bf k}$-sum, we obtain  
\begin{align}
\chi^{+-,(0)}_{AA}({\bf q},\omega) &= \frac{1}{2\Delta} \left[ 1 + \frac{\omega}{2\Delta}
- \frac{4t_1^2}{\Delta^2} + \frac{t_2^2 (1-\cos q_x) - 2t_2^3 (1-\cos q_x \cos q_y)
}{\Delta^2}  \right] \notag \\
\chi^{+-,(0)}_{BB}({\bf q},\omega) &= \frac{1}{2\Delta} \left[ 1 - \frac{\omega}{2\Delta}
- \frac{4t_1^2}{\Delta^2} + \frac{t_2^2 (1-\cos q_x) - 2t_2^3 (1-\cos q_x \cos q_y)
}{\Delta^2}  \right]
\notag \\
\chi^{+-,(0)}_{AB}({\bf q},\omega) &=  \chi^{+-,(0)}_{BA}({\bf q},\omega)  = - \frac{t^2_1}{\Delta^3} \cos \frac{q_x}{2} \cos \frac{q_y}{2}.
\end{align}
Here, we have taken low-temperature limit and analytical continuation to extract the real part of the matrix elements. The elements of $\chi^{-+,(0)}$ can be obtained similarly and we arrive at 
\begin{align}
\chi^{-+,(0)}_{\mu \nu}({\bf q},\omega) = \chi^{+-,(0)}_{\mu \nu}({\bf q},- \omega).
\end{align}
Consequently, we have
\begin{align}
\mathbbm{1}- U \chi^{\kappa,(0)}({\bf q},\omega) = \frac{2t^2_1}{\Delta^2} 
\begin{pmatrix}
1 - \kappa \dfrac{\omega}{2 m J_1}-\dfrac{J_2}{2J_1} \gamma'_{q_x} - \dfrac{J_3}{J_1} \gamma''_{\bf q} & \gamma_{\bf q}\\
\gamma_{\bf q} & 1 + \kappa \dfrac{\omega}{2 m J_1}-\dfrac{J_2}{2J_1}  \gamma'_{q_x} - \dfrac{J_3}{J_1} \gamma''_{\bf q}
\end{pmatrix},
\end{align}
where $m \approx \left(1-2t_1^2/\Delta^2 \right)$ is the magnetization, $J_i = {4t^2_i}/{U}$, $\gamma_{\bf q} =  \cos \dfrac{q_x}{2} \cos \dfrac{q_y}{2}$, $\gamma'_{q_{x/y}} = \sin^2 \dfrac{q_{x/y}}{2}$, $ \gamma''_{\bf q}=\left( 1 - \cos q_x \cos q_y \right)$, and the $\kappa = + 1$ ($-1$) refer to positive (negative) chirality magnons. Then the magnon energies obtained from the pole of $(\mathbbm{1}- U \chi^{\kappa,(0)}({\bf q},\omega_{\bf q}))$ read
\begin{align}
\frac{\omega_{{\bf q},\kappa}}{2 m J_1} =   \sqrt{\left( 1 - \dfrac{J_2}{2J_1}\left( \gamma'_{q_x} + \gamma'_{q_y}\right) - \frac{J_3}{J_1}\gamma''_{\bf q}  \right)^2 - \gamma^2_{\bf q} } \; - \kappa \; \dfrac{J_2}{2J_1}\left( \gamma'_{q_x} - \gamma'_{q_y}\right).
\end{align}

\section{Spin-wave dispersion in a Heisenberg model}
\label{sec:sw_heisenberg}
For completeness, we show here that a spin-wave dispersion similar to that obtained in the Hubbard model in the strong-coupling limit can also be obtained in the spin Hamiltonian
\begin{align}
{\cal H}_s = \sum_{i < j}  J_{ij} {\bf S}_i \cdot {\bf S}_j,
\end{align}
where $J_{ij} > 0$ are the exchange couplings. We include first-, second-, and third-neighbor exchange processes, parametrized by three exchange constants, $J_1$ , $J_2$ , and $J_3$ , respectively. We write the spin operators in a Holstein-Primakoff
representation retaining terms up to second order in the magnon operators as
\begin{align}
S_i^+ &= \sqrt{2S} a_i, \qquad S_i^- = \sqrt{2S} a_i^\dagger, \qquad S_i^z = S - a_i^\dagger a_i, \notag \\
S_i^+ &= \sqrt{2S} b_i^\dagger, \qquad  S_i^- = \sqrt{2S} b_i, \qquad  S_i^z = -S + b_i^\dagger b_i. 
\end{align}
Here, $a_i$ ($b_i$) is the bosonic annihilation operator for the $A$ ($B$) sublattice.
After Fourier transformation, the spin Hamiltonian can be written in terms of magnon operators as the following:
\begin{align}
{\cal H}_s =\begin{pmatrix}
    a_{\bf q}^\dagger & b_{-{\bf q}}
\end{pmatrix}
\begin{pmatrix}
{\cal A}_{\bf q} & {\cal C}_{\bf q} \\
{\cal C}^*_{\bf q} & {\cal B}_{\bf q}
\end{pmatrix}
\begin{pmatrix}
    a_{\bf q} \\ b_{-{\bf q}}^\dagger
\end{pmatrix},
\end{align}
where ${\cal A}_{\bf q} = 4 S \left( J_1 - J_2 \gamma'_{q_y} - J_3 \gamma''_{\bf q} \right) $, ${\cal B}_{\bf q} = 4 S \left( J_1 - J_2 \gamma'_{q_x} - J_3 \gamma''_{\bf q} \right)$, ${\cal C}_{\bf q} = 4 J_1 S \, \gamma_{\bf q}$, with $\gamma_{\bf q}$, $\gamma'_{q_{x/y}}$, and $\gamma''_{\bf q}$ being defined in the main text. To obtain magnon frequencies, we carry out a Bogolioubov transformation $\alpha_{\bf q} = u_{\bf q} a_{\bf q} -v_{\bf q} b_{-{\bf q}}^\dagger$ and $\beta_{\bf q} =u_{\bf q} b_{-{\bf q}} -v_{\bf q} a_{\bf q}^\dagger$ with $\alpha_{\bf q}$ and $\beta_{\bf q}$ being the annihilation operators for the two magnon branches. The Heisenberg equations of motion $i \dot{\alpha}_{\bf q} = [\alpha_{\bf q}, {\cal H}_s]$ and $i \dot{\beta}_{\bf q} = [\beta_{\bf q}, {\cal H}_s]$ yield the eigenequations of the Bogoliubov wave function. Solving them, we obtain the dispersion relation
\begin{align}
\frac{\omega_{{\bf q},\kappa}}{4 S J_1} =   \sqrt{\left( 1 - \dfrac{J_2}{2J_1}\left( \gamma'_{q_x} + \gamma'_{q_y}\right) - \frac{J_3}{J_1}\gamma''_{\bf q}  \right)^2 - \gamma^2_{\bf q} } \; - \kappa \; \dfrac{J_2}{2J_1}\left( \gamma'_{q_x} - \gamma'_{q_y}\right),
\end{align}
with $\kappa$ being the magnon chirality. Replacing $2S$ by the magnetization $m$ in the above expression recovers the spin-wave dispersion of the Hubbard model in the strong-coupling limit.
\end{widetext}

\bibliography{refs}{}

@article{roig2024minimal,
  title = {Minimal models for altermagnetism},
  author = {Roig, Merc\`e and Kreisel, Andreas and Yu, Yue and Andersen, Brian M. and Agterberg, Daniel F.},
  journal = {Phys. Rev. B},
  volume = {110},
  issue = {14},
  pages = {144412},
  numpages = {20},
  year = {2024},
  month = {Oct},
  publisher = {American Physical Society},
  doi = {10.1103/PhysRevB.110.144412}}

@book{Auerbach1994, 
 title={Interacting Electrons and Quantum Magnetism}, 
 ISBN={9781461208693}, 
 ISSN={0938-037X}, 
 url={http://dx.doi.org/10.1007/978-1-4612-0869-3}, 
 DOI={10.1007/978-1-4612-0869-3}, 
series = {Graduate Texts in Contemporary Physics},
 publisher={Springer New York}, 
 author={Auerbach, Assa}, 
 year={1994} }

@article{smejkal2023chiral,
  title = {Chiral Magnons in Altermagnetic {RuO$_2$}},
  author = {\ifmmode \check{S}\else \v{S}\fi{}mejkal, Libor and Marmodoro, Alberto and Ahn, Kyo-Hoon and Gonz\'alez-Hern\'andez, Rafael and Turek, Ilja and Mankovsky, Sergiy and Ebert, Hubert and D'Souza, Sunil W. and \ifmmode \check{S}\else \v{S}\fi{}ipr, Ond\ifmmode \check{r}\else \v{r}\fi{}ej and Sinova, Jairo and Jungwirth, Tom\'a\ifmmode \check{s}\else \v{s}\fi{}},
  journal = {Phys. Rev. Lett.},
  volume = {131},
  issue = {25},
  pages = {256703},
  numpages = {6},
  year = {2023},
  month = {Dec},
  publisher = {American Physical Society},
  doi = {10.1103/PhysRevLett.131.256703},
  url = {https://link.aps.org/doi/10.1103/PhysRevLett.131.256703}
}

@article{zhang2025chiral,
  title = {Chiral magnon splitting in altermagnetic {CrSb} from first principles},
  author = {Zhang, Yi-Fan and Ni, Xiao-Sheng and Chen, Ke and Cao, Kun},
  journal = {Phys. Rev. B},
  volume = {111},
  issue = {17},
  pages = {174451},
  numpages = {7},
  year = {2025},
  month = {May},
  publisher = {American Physical Society},
  doi = {10.1103/PhysRevB.111.174451},
  url = {https://link.aps.org/doi/10.1103/PhysRevB.111.174451}
}

@article{liu2024chiral,
  title = {Chiral Split Magnon in Altermagnetic {MnTe}},
  author = {Liu, Zheyuan and Ozeki, Makoto and Asai, Shinichiro and Itoh, Shinichi and Masuda, Takatsugu},
  journal = {Phys. Rev. Lett.},
  volume = {133},
  issue = {15},
  pages = {156702},
  numpages = {6},
  year = {2024},
  month = {Oct},
  publisher = {American Physical Society},
  doi = {10.1103/PhysRevLett.133.156702},
  url = {https://link.aps.org/doi/10.1103/PhysRevLett.133.156702}
}

@article{gonzalez2026chiral, 
title={Topological Chiral Magnons in the Altermagnet {AgF$_2$}
                    Monolayer}, volume={43}, ISSN={1741-3540}, url={http://dx.doi.org/10.1088/0256-307x/43/2/020705}, DOI={10.1088/0256-307x/43/2/020705}, number={2}, journal={Chin. Phys. Lett.}, publisher={IOP Publishing}, author={González, J. W. and Gallardo, R. A. and Vidal-Silva, N. and León, A. M.}, year={2026}, month=feb, pages={020705} }

@article{consoli2025altermagnetism,
  title = {$\mathrm{SU}(N)$ Altermagnetism: Lattice Models, Magnon Modes, and Flavor-Split Bands},
  author = {C\^onsoli, Pedro M. and Vojta, Matthias},
  journal = {Phys. Rev. Lett.},
  volume = {134},
  issue = {19},
  pages = {196701},
  numpages = {6},
  year = {2025},
  month = {May},
  publisher = {American Physical Society},
  doi = {10.1103/PhysRevLett.134.196701},
  url = {https://link.aps.org/doi/10.1103/PhysRevLett.134.196701}
}

@article{costa2025giant,
	title={{Giant spatial anisotropy of magnon Landau damping in altermagnets}},
	author={António T. Costa and João C. G. Henriques and Joaquín Fernández-Rossier},
	journal={SciPost Phys.},
	volume={18},
	number={4},
	pages={125},
	year={2025},
	publisher={SciPost},
	doi={10.21468/SciPostPhys.18.4.125},
	url={https://scipost.org/10.21468/SciPostPhys.18.4.125},
}

@article{hoyer2025altermagnetic,
  title = {Altermagnetic splitting of magnons in hematite {$\alpha$-{Fe$_2$O$_3$}}},
  author = {Hoyer, Rhea and Stavropoulos, P. Peter and Razpopov, Aleksandar and Valent\'{\i}, Roser and \ifmmode \check{S}\else \v{S}\fi{}mejkal, Libor and Mook, Alexander},
  journal = {Phys. Rev. B},
  volume = {112},
  issue = {6},
  pages = {064425},
  numpages = {22},
  year = {2025},
  month = {Aug},
  publisher = {American Physical Society},
  doi = {10.1103/fgc1-5blp},
  url = {https://link.aps.org/doi/10.1103/fgc1-5blp}
}

@article{sun2025observation,
  title = {Observation of Chiral Magnon Band Splitting in Altermagnetic Hematite},
  author = {Sun, Qiyang and Guo, Jiasen and Wang, Dan and Abernathy, Douglas L. and Tian, Wei and Li, Chen},
  journal = {Phys. Rev. Lett.},
  volume = {135},
  issue = {18},
  pages = {186703},
  numpages = {6},
  year = {2025},
  month = {Oct},
  publisher = {American Physical Society},
  doi = {10.1103/7yhz-jptc},
  url = {https://link.aps.org/doi/10.1103/7yhz-jptc}
}

@article{maier2023weak,
  title = {Weak-coupling theory of neutron scattering as a probe of altermagnetism},
  author = {Maier, Thomas A. and Okamoto, Satoshi},
  journal = {Phys. Rev. B},
  volume = {108},
  issue = {10},
  pages = {L100402},
  numpages = {5},
  year = {2023},
  month = {Sep},
  publisher = {American Physical Society},
  doi = {10.1103/PhysRevB.108.L100402},
  url = {https://link.aps.org/doi/10.1103/PhysRevB.108.L100402}
}

@article{das2024realizing,
  title = {Realizing Altermagnetism in {Fermi}-{Hubbard} model with Ultracold Atoms},
  author = {Das, Purnendu and Leeb, Valentin and Knolle, Johannes and Knap, Michael},
  journal = {Phys. Rev. Lett.},
  volume = {132},
  issue = {26},
  pages = {263402},
  numpages = {7},
  year = {2024},
  month = {Jun},
  publisher = {American Physical Society},
  doi = {10.1103/PhysRevLett.132.263402},
  url = {https://link.aps.org/doi/10.1103/PhysRevLett.132.263402}
}

@article{smejkal2022emerging,
  title = {Emerging Research Landscape of Altermagnetism},
  author = {\ifmmode \check{S}\else \v{S}\fi{}mejkal, Libor and Sinova, Jairo and Jungwirth, Tomas},
  journal = {Phys. Rev. X},
  volume = {12},
  issue = {4},
  pages = {040501},
  numpages = {27},
  year = {2022},
  month = {Dec},
  publisher = {American Physical Society},
  doi = {10.1103/PhysRevX.12.040501},
  url = {https://link.aps.org/doi/10.1103/PhysRevX.12.040501}
}

@article{smejkal2022beyond,
  title = {Beyond Conventional Ferromagnetism and Antiferromagnetism: A Phase with Nonrelativistic Spin and Crystal Rotation Symmetry},
  author = {\ifmmode \check{S}\else \v{S}\fi{}mejkal, Libor and Sinova, Jairo and Jungwirth, Tomas},
  journal = {Phys. Rev. X},
  volume = {12},
  issue = {3},
  pages = {031042},
  numpages = {16},
  year = {2022},
  month = {Sep},
  publisher = {American Physical Society},
  doi = {10.1103/PhysRevX.12.031042},
  url = {https://link.aps.org/doi/10.1103/PhysRevX.12.031042}
}

@article{xiao2024spin,
  title = {Spin Space Groups: Full Classification and Applications},
  author = {Xiao, Zhenyu and Zhao, Jianzhou and Li, Yanqi and Shindou, Ryuichi and Song, Zhi-Da},
  journal = {Phys. Rev. X},
  volume = {14},
  issue = {3},
  pages = {031037},
  numpages = {33},
  year = {2024},
  month = {Aug},
  publisher = {American Physical Society},
  doi = {10.1103/PhysRevX.14.031037},
  url = {https://link.aps.org/doi/10.1103/PhysRevX.14.031037}
}

@article{jiang2024enumeration,
  title = {Enumeration of Spin-Space Groups: Toward a Complete Description of Symmetries of Magnetic Orders},
  author = {Jiang, Yi and Song, Ziyin and Zhu, Tiannian and Fang, Zhong and Weng, Hongming and Liu, Zheng-Xin and Yang, Jian and Fang, Chen},
  journal = {Phys. Rev. X},
  volume = {14},
  issue = {3},
  pages = {031039},
  numpages = {25},
  year = {2024},
  month = {Aug},
  publisher = {American Physical Society},
  doi = {10.1103/PhysRevX.14.031039},
  url = {https://link.aps.org/doi/10.1103/PhysRevX.14.031039}
}

@article{chen2024enumeration,
  title = {Enumeration and Representation Theory of Spin Space Groups},
  author = {Chen, Xiaobing and Ren, Jun and Zhu, Yanzhou and Yu, Yutong and Zhang, Ao and Liu, Pengfei and Li, Jiayu and Liu, Yuntian and Li, Caiheng and Liu, Qihang},
  journal = {Phys. Rev. X},
  volume = {14},
  issue = {3},
  pages = {031038},
  numpages = {33},
  year = {2024},
  month = {Aug},
  publisher = {American Physical Society},
  doi = {10.1103/PhysRevX.14.031038},
  url = {https://link.aps.org/doi/10.1103/PhysRevX.14.031038}
}

@article{gao2025ai,
   title={{AI}-accelerated discovery of altermagnetic materials},
   volume={12},
   ISSN={2053-714X},
   url = {https://doi.org/10.1093/nsr/nwaf066},
   DOI={10.1093/nsr/nwaf066},
   number={4},
   journal={Natl. Sci. Rev.},
   pages = {nwaf066},
   publisher={Oxford University Press (OUP)},
   author={Gao, Ze-Feng and Qu, Shuai and Zeng, Bocheng and Liu, Yang and Wen, Ji-Rong and Sun, Hao and Guo, Peng-Jie and Lu, Zhong-Yi},
   year={2025},
   month=feb }

@article{shuai2026extremely,
title = {Extremely strong spin-orbit coupling effect in light-element altermagnetic materials},
journal = {Front. Phys.},
volume = {21},
pages = {045203},
year = {2026},
issn = {2095-0462},
doi={10.15302/frontphys.2026.045203},
url = {https://journal.hep.com.cn/fop/EN/10.15302/frontphys.2026.045203},
  author={Qu, Shuai and Ouyang, Zhen-Feng and Gao, Ze-Feng and Sun, Hao and Liu, Kai and Guo, Peng-Jie and Lu, Zhong-Yi},
}

@misc{tamang2024newly,
      title={Newly discovered magnetic phase: A brief review on Altermagnets}, 
      author={R. Tamang and Shivraj Gurung and D. P. Rai and Samy Brahimi and Samir Lounis},
      eprint={2412.05377},
      archivePrefix={arXiv},
      url={https://arxiv.org/abs/2412.05377}, 
}

@article{fedchenko2024observation,
author = {Fedchenko, Olena and others},
title = {Observation of time-reversal symmetry breaking in the band structure of altermagnetic {RuO$_2$}},
journal = {Sci. Adv.},
volume = {10},
number = {5},
pages = {eadj4883},
year = {2024},
doi = {10.1126/sciadv.adj4883},
URL = {https://www.science.org/doi/abs/10.1126/sciadv.adj4883},
}

@article{ding2024large,
  title = {Large Band Splitting in $g$-Wave Altermagnet {CrSb}},
  author = {Ding, Jianyang and others},
  journal = {Phys. Rev. Lett.},
  volume = {133},
  issue = {20},
  pages = {206401},
  numpages = {7},
  year = {2024},
  month = {Nov},
  publisher = {American Physical Society},
  doi = {10.1103/PhysRevLett.133.206401},
  url = {https://link.aps.org/doi/10.1103/PhysRevLett.133.206401}
}

@article{attias2024intrinsic,
  title = {Intrinsic anomalous Hall effect in altermagnets},
  author = {Attias, Lotan and Levchenko, Alex and Khodas, Maxim},
  journal = {Phys. Rev. B},
  volume = {110},
  issue = {9},
  pages = {094425},
  numpages = {10},
  year = {2024},
  month = {Sep},
  publisher = {American Physical Society},
  doi = {10.1103/PhysRevB.110.094425},
  url = {https://link.aps.org/doi/10.1103/PhysRevB.110.094425}
}

@article{cui2023efficient,
  title = {Efficient spin Seebeck and spin Nernst effects of magnons in altermagnets},
  author = {Cui, Qirui and Zeng, Bowen and Cui, Ping and Yu, Tao and Yang, Hongxin},
  journal = {Phys. Rev. B},
  volume = {108},
  issue = {18},
  pages = {180401},
  numpages = {7},
  year = {2023},
  month = {Nov},
  publisher = {American Physical Society},
  doi = {10.1103/PhysRevB.108.L180401},
  url = {https://link.aps.org/doi/10.1103/PhysRevB.108.L180401}
}

@article{zhou2024crystal,
  title = {Crystal Thermal Transport in Altermagnetic {RuO$_2$}},
  author = {Zhou, Xiaodong and Feng, Wanxiang and Zhang, Run-Wu and \ifmmode \check{S}\else \v{S}\fi{}mejkal, Libor and Sinova, Jairo and Mokrousov, Yuriy and Yao, Yugui},
  journal = {Phys. Rev. Lett.},
  volume = {132},
  issue = {5},
  pages = {056701},
  numpages = {7},
  year = {2024},
  month = {Jan},
  publisher = {American Physical Society},
  doi = {10.1103/PhysRevLett.132.056701},
  url = {https://link.aps.org/doi/10.1103/PhysRevLett.132.056701}
}

@article{solovyev1997magneto-optical,
  title = {Magneto-optical effect in the weak ferromagnets {$\mathrm{{La}{M}{O}_{3}}$ ({M} = {Cr}, {Mn}, and {Fe})}},
  author = {Solovyev, I. V.},
  journal = {Phys. Rev. B},
  volume = {55},
  issue = {13},
  pages = {8060--8063},
  numpages = {0},
  year = {1997},
  month = {Apr},
  publisher = {American Physical Society},
  doi = {10.1103/PhysRevB.55.8060},
  url = {https://link.aps.org/doi/10.1103/PhysRevB.55.8060}
}

@article{vakili2025spin,
  title = {Spin-Transfer Torque in Altermagnets with Magnetic Textures},
  author = {Vakili, Hamed and Schwartz, Edward and Kovalev, Alexey A.},
  journal = {Phys. Rev. Lett.},
  volume = {134},
  issue = {17},
  pages = {176401},
  numpages = {6},
  year = {2025},
  month = {Apr},
  publisher = {American Physical Society},
  doi = {10.1103/PhysRevLett.134.176401},
  url = {https://link.aps.org/doi/10.1103/PhysRevLett.134.176401}
}

@article{yershov2024fluctuation,
  title = {Fluctuation-induced piezomagnetism in local moment altermagnets},
  author = {Yershov, Kostiantyn V. and Kravchuk, Volodymyr P. and Daghofer, Maria and van den Brink, Jeroen},
  journal = {Phys. Rev. B},
  volume = {110},
  issue = {14},
  pages = {144421},
  numpages = {11},
  year = {2024},
  month = {Oct},
  publisher = {American Physical Society},
  doi = {10.1103/PhysRevB.110.144421},
  url = {https://link.aps.org/doi/10.1103/PhysRevB.110.144421}
}

@article{liu2025unconventional,
author = {Liu, Yuntian and Shrestha, Reshna and Denisov, Konstantin and Ayala, Denzel and van Schilfgaarde, Mark and Nie, Wanyi and Žutić, Igor},
title = {Unconventional Spintronics from Chiral Perovskites},
journal = {Advanced Functional Materials},
volume = {35},
number = {52},
pages = {e09127},
doi = {https://doi.org/10.1002/adfm.202509127},
url = {https://advanced.onlinelibrary.wiley.com/doi/abs/10.1002/adfm.202509127},
year = {2025}
}

@misc{spaldin2026why,
      title={Why are there so few non-altermagnetic antiferromagnets?}, 
      author={Nicola A. Spaldin and Sang-Wook Cheong and Sinead Griffin},
      eprint={2602.17181},
      archivePrefix={arXiv},
      url={https://arxiv.org/abs/2602.17181}, 
}

@article{feng2022anomalous,
  author  = {Z. Feng and others},
  title   = {An anomalous Hall effect in altermagnetic ruthenium dioxide},
  journal = {Nat. Electron.},
  volume  = {5},
  number  = {11},
  pages   = {735--743},
  doi     = {10.1038/s41928-022-00841-z},
   year   = {2022},
}

@article{smejkal2020crystal,
   title={Crystal time-reversal symmetry breaking and spontaneous Hall effect in collinear antiferromagnets},
   volume={6},
   ISSN={2375-2548},
   url={http://dx.doi.org/10.1126/sciadv.aaz8809},
   number={23},
   pages = {eaaz8809},
   journal={Sci. Adv.},
   publisher={American Association for the Advancement of Science (AAAS)},
   author={Šmejkal, Libor and González-Hernández, Rafael and Jungwirth, T. and Sinova, J.},
   year={2020},
   month = {June} }

@article{mazin2021prediction,
   title={Prediction of unconventional magnetism in doped {FeSb$_2$}},
   volume={118},
   ISSN={1091-6490},
   url={http://dx.doi.org/10.1073/pnas.2108924118},
   number={42},
   journal={Proc. Natl. Acad. Sci. U.S.A.},
   publisher={Proceedings of the National Academy of Sciences},
   author={Mazin, Igor I. and Koepernik, Klaus and Johannes, Michelle D. and González-Hernández, Rafael and Šmejkal, Libor},
   year={2021},
   month=Oct}

@article{ahn2019antiferromagnetism,
  title = {Antiferromagnetism in {RuO$_2$} as $d$-wave Pomeranchuk instability},
  author = {Ahn, Kyo-Hoon and Hariki, Atsushi and Lee, Kwan-Woo and Kune\ifmmode \check{s}\else \v{s}\fi{}, Jan},
  journal = {Phys. Rev. B},
  volume = {99},
  issue = {18},
  pages = {184432},
  numpages = {5},
  year = {2019},
  month = {May},
  publisher = {American Physical Society},
  doi = {10.1103/PhysRevB.99.184432},
  url = {https://link.aps.org/doi/10.1103/PhysRevB.99.184432}
  }

@article{yuan2020giant,
  title = {Giant momentum-dependent spin splitting in centrosymmetric low-{$Z$} antiferromagnets},
  author = {Yuan, Lin-Ding and Wang, Zhi and Luo, Jun-Wei and Rashba, Emmanuel I. and Zunger, Alex},
  journal = {Phys. Rev. B},
  volume = {102},
  issue = {1},
  pages = {014422},
  numpages = {13},
  year = {2020},
  month = {Jul},
  publisher = {American Physical Society},
  doi = {10.1103/PhysRevB.102.014422},
  url = {https://link.aps.org/doi/10.1103/PhysRevB.102.014422}
}

@article{yuan2021prediction,
  title = {Prediction of low-{$Z$} collinear and noncollinear antiferromagnetic compounds having momentum-dependent spin splitting even without spin-orbit coupling},
  author = {Yuan, Lin-Ding and Wang, Zhi and Luo, Jun-Wei and Zunger, Alex},
  journal = {Phys. Rev. Mater.},
  volume = {5},
  issue = {1},
  pages = {014409},
  numpages = {24},
  year = {2021},
  month = {Jan},
  publisher = {American Physical Society},
  doi = {10.1103/PhysRevMaterials.5.014409},
  url = {https://link.aps.org/doi/10.1103/PhysRevMaterials.5.014409}
}

@article{egorov2021colossal,
author = {Egorov, Sergei and Robert, Evarestov},
year = {2021},
month = {03},
pages = {2363-2369},
title = {Colossal Spin Splitting in the Monolayer of the Collinear Antiferromagnet {MnF$_2$}},
volume = {12},
journal = {J. Phys. Chem. Lett.},
doi = {10.1021/acs.jpclett.1c00282}
}

@article{zhu2025design,
  title = {Design of Altermagnetic Models from Spin Clusters},
  author = {Zhu, Xingchuan and Huo, Xingmin and Feng, Shiping and Zhang, Song-Bo and Yang, Shengyuan A. and Guo, Huaiming},
  journal = {Phys. Rev. Lett.},
  volume = {134},
  issue = {16},
  pages = {166701},
  numpages = {6},
  year = {2025},
  month = {Apr},
  publisher = {American Physical Society},
  doi = {10.1103/PhysRevLett.134.166701},
  url = {https://link.aps.org/doi/10.1103/PhysRevLett.134.166701}
}

@article{vijayvargia2025altermagnets,
  title = {Altermagnets with Topological Order in Kitaev Bilayers},
  author = {Vijayvargia, Aayush and Day-Roberts, Ezra and Botana, Antia S. and Erten, Onur},
  journal = {Phys. Rev. Lett.},
  volume = {135},
  issue = {16},
  pages = {166701},
  numpages = {7},
  year = {2025},
  month = {Oct},
  publisher = {American Physical Society},
  doi = {10.1103/km2j-3zy2},
  url = {https://link.aps.org/doi/10.1103/km2j-3zy2}
}

@misc{wiedmann2025quantum,
      title={Quantum effects in the magnon spectrum of 2{D} altermagnets via continuous similarity transformations}, 
      author={Raymond Wiedmann and Dag-Björn Hering and Vanessa Sulaiman and Matthias R. Walther and Kai P. Schmidt and Götz S. Uhrig},
      eprint={2511.03528},
      archivePrefix={arXiv},
      url={https://arxiv.org/abs/2511.03528}, 
      }

@article{daghofer2026altermagnetic,
  title = {Altermagnetic Polarons: The Fate of Altermagnetic Band Splittings at Strong Coupling},
  author = {Daghofer, Maria and Wohlfeld, Krzysztof and van den Brink, Jeroen},
  journal = {Phys. Rev. Lett.},
  volume = {136},
  issue = {14},
  pages = {146502},
  numpages = {7},
  year = {2026},
  month = {Apr},
  publisher = {American Physical Society},
  doi = {10.1103/261t-cs82},
  url = {https://link.aps.org/doi/10.1103/261t-cs82}
}

@article{lieb1999ground,
  title = {Ground State Properties of a Fully Frustrated Quantum Spin System},
  author = {Lieb, Elliott H. and Schupp, Peter},
  journal = {Phys. Rev. Lett.},
  volume = {83},
  issue = {25},
  pages = {5362--5365},
  numpages = {0},
  year = {1999},
  month = {Dec},
  publisher = {American Physical Society},
  doi = {10.1103/PhysRevLett.83.5362},
  url = {https://link.aps.org/doi/10.1103/PhysRevLett.83.5362}
}

@article{palmer2000order,
  title = {Order induced by dipolar interactions in a geometrically frustrated antiferromagnet},
  author = {Palmer, S. E. and Chalker, J. T.},
  journal = {Phys. Rev. B},
  volume = {62},
  issue = {1},
  pages = {488--492},
  numpages = {0},
  year = {2000},
  month = {Jul},
  publisher = {American Physical Society},
  doi = {10.1103/PhysRevB.62.488},
  url = {https://link.aps.org/doi/10.1103/PhysRevB.62.488}
}

@article{palmer2001quantum,
  title = {Quantum disorder in the two-dimensional pyrochlore Heisenberg antiferromagnet},
  author = {Palmer, S. E. and Chalker, J. T.},
  journal = {Phys. Rev. B},
  volume = {64},
  issue = {9},
  pages = {094412},
  numpages = {6},
  year = {2001},
  month = {Aug},
  publisher = {American Physical Society},
  doi = {10.1103/PhysRevB.64.094412},
  url = {https://link.aps.org/doi/10.1103/PhysRevB.64.094412}
}

@article{starykh2002spinons,
  title = {Spinons in a Crossed-Chains Model of a 2{D} Spin Liquid},
  author = {Starykh, Oleg A. and Singh, Rajiv R. P. and Levine, Gregory C.},
  journal = {Phys. Rev. Lett.},
  volume = {88},
  issue = {16},
  pages = {167203},
  numpages = {4},
  year = {2002},
  month = {Apr},
  publisher = {American Physical Society},
  doi = {10.1103/PhysRevLett.88.167203},
  url = {https://link.aps.org/doi/10.1103/PhysRevLett.88.167203}
}

@article{canals2002from,
  title = {From the square lattice to the checkerboard lattice: Spin-wave and large-$n$ limit analysis},
  author = {Canals, Benjamin},
  journal = {Phys. Rev. B},
  volume = {65},
  issue = {18},
  pages = {184408},
  numpages = {8},
  year = {2002},
  month = {Apr},
  publisher = {American Physical Society},
  doi = {10.1103/PhysRevB.65.184408},
  url = {https://link.aps.org/doi/10.1103/PhysRevB.65.184408}
}

@article{berg2003singlet,
  title = {Singlet Excitations in Pyrochlore: A Study of Quantum Frustration},
  author = {Berg, Erez and Altman, Ehud and Auerbach, Assa},
  journal = {Phys. Rev. Lett.},
  volume = {90},
  issue = {14},
  pages = {147204},
  numpages = {4},
  year = {2003},
  month = {Apr},
  publisher = {American Physical Society},
  doi = {10.1103/PhysRevLett.90.147204},
  url = {https://link.aps.org/doi/10.1103/PhysRevLett.90.147204}
}

@article{Tchernyshyov2003bond,
  title = {Bond order from disorder in the planar pyrochlore magnet},
  author = {Tchernyshyov, O. and Starykh, O. A. and Moessner, R. and Abanov, A. G.},
  journal = {Phys. Rev. B},
  volume = {68},
  issue = {14},
  pages = {144422},
  numpages = {17},
  year = {2003},
  month = {Oct},
  publisher = {American Physical Society},
  doi = {10.1103/PhysRevB.68.144422},
  url = {https://link.aps.org/doi/10.1103/PhysRevB.68.144422}
}

@article{bernier2004planar,
  title = {Planar pyrochlore antiferromagnet: A large-$N$ analysis},
  author = {Bernier, Jean-S\'ebastien and Chung, Chung-Hou and Kim, Yong Baek and Sachdev, Subir},
  journal = {Phys. Rev. B},
  volume = {69},
  issue = {21},
  pages = {214427},
  numpages = {7},
  year = {2004},
  month = {Jun},
  publisher = {American Physical Society},
  doi = {10.1103/PhysRevB.69.214427},
  url = {https://link.aps.org/doi/10.1103/PhysRevB.69.214427}
}

@article{moessner2004planar, title={Planar Pyrochlore, Quantum Ice and Sliding Ice}, volume={116}, ISSN={1572-9613}, url={http://dx.doi.org/10.1023/b:joss.0000037247.54022.62}, DOI={10.1023/b:joss.0000037247.54022.62}, number={1-4}, journal={J. Stat. Phys.}, publisher={Springer Science and Business Media LLC}, author={Moessner, R. and Tchernyshyov, Oleg and Sondhi, S. L.}, year={2004}, month=Aug, pages={755–772} }

@misc{jungwirth2025altermagneticspintronics,
      title={Altermagnetic spintronics}, 
      author={T. Jungwirth and J. Sinova and P. Wadley and D. Kriegner and H. Reichlova and F. Krizek and H. Ohno and L. Smejkal},
      eprint={2508.09748},
      archivePrefix={arXiv},
      url={https://arxiv.org/abs/2508.09748}, 
}

@article{smejkal2022anomalous,
  title={Anomalous hall antiferromagnets},
  author={{\v{S}}mejkal, Libor and MacDonald, Allan H and Sinova, Jairo and Nakatsuji, Satoru and Jungwirth, Tomas},
  journal={Nat. Rev. Mater.},
  volume={7},
  number={6},
  pages={482--496},
  year={2022},
  publisher={Nature Publishing Group UK London},
  doi={10.1038/s41578-022-00430-3},
 url={https://doi.org/10.1038/s41578-022-00430-3} 
}

@misc{li2025enhancement,
      title={Enhancement of $d$-wave pairing in Strongly Correlated Altermagnet}, 
      author={Jianyu Li and Ji Liu and Xiaosen Yang and Ho-Kin Tang},
      eprint={2505.12342},
      archivePrefix={arXiv},
      url={https://arxiv.org/abs/2505.12342}, 
}

@article{ma2024altermagnetic,
author={Ma, Hai-Yang and Jia, Jin-Feng},
title = {Altermagnetic topological insulator with ${\mathcal C}$-paired spin-valley locking},
journal= {Quantum Front.},
year={2024},
month={Nov},
volume={3},
number={1},
pages={22},
doi={10.1007/s44214-024-00070-4},
url={https://doi.org/10.1007/s44214-024-00070-4}
}

@article{he2025altermagnetism,
  title = {Altermagnetism and beyond in the $t\text{\ensuremath{-}}{t}^{\ensuremath{'}}\text{\ensuremath{-}}\ensuremath{\delta}$ {Fermi}-{Hubbard} model},
  author = {He, Saisai and Zhao, Jize and Luo, Hong-Gang and Hu, Shijie},
  journal = {Phys. Rev. B},
  volume = {112},
  issue = {3},
  pages = {035108},
  numpages = {13},
  year = {2025},
  month = {Jul},
  publisher = {American Physical Society},
  doi = {10.1103/4mv8-tb66},
  url = {https://link.aps.org/doi/10.1103/4mv8-tb66}
}

@article{bose2024altermagnetism,
  title = {Altermagnetism and superconductivity in a multiorbital {$t$}-{J} model},
  author = {Bose, Anjishnu and Vadnais, Samuel and Paramekanti, Arun},
  journal = {Phys. Rev. B},
  volume = {110},
  issue = {20},
  pages = {205120},
  numpages = {15},
  year = {2024},
  month = {Nov},
  publisher = {American Physical Society},
  doi = {10.1103/PhysRevB.110.205120},
  url = {https://link.aps.org/doi/10.1103/PhysRevB.110.205120}
}

@article{che2025engineering, 
  title = {Engineering Altermagnetic States in Two-Dimensional Square Tessellations},
  author = {Che, Yixuan and Lv, Haifeng and Wu, Xiaojun and Yang, Jinlong},
  journal = {Phys. Rev. Lett.},
  volume = {135},
  issue = {3},
  pages = {036701},
  numpages = {7},
  year = {2025},
  month = {Jul},
  publisher = {American Physical Society},
  doi = {10.1103/v38b-5by1},
  url = {https://link.aps.org/doi/10.1103/v38b-5by1}}

@article{wang2026spin,
  title = {Spin-biased quantum spin Hall effect in altermagnetic {Lieb} lattice},
  author = {Wang, Qianjun and Wu, Ruqian and Hu, Jun},
  journal = {Phys. Rev. B},
  volume = {113},
  issue = {16},
  pages = {161101},
  numpages = {8},
  year = {2026},
  month = {Apr},
  publisher = {American Physical Society},
  doi = {10.1103/kqwx-v6jv},
  url = {https://link.aps.org/doi/10.1103/kqwx-v6jv}
}

@article{kaushal2025altermagnetism,
  title = {Altermagnetism in Modified {Lieb} lattice {Hubbard} model},
  author = {Kaushal, Nitin and Franz, Marcel},
  journal = {Phys. Rev. Lett.},
  volume = {135},
  issue = {15},
  pages = {156502},
  numpages = {8},
  year = {2025},
  month = {Oct},
  publisher = {American Physical Society},
  doi = {10.1103/s31h-hk2v},
  url = {https://link.aps.org/doi/10.1103/s31h-hk2v}
}

@misc{biswas2026altermagnetic,
      title={Altermagnetic phases and phase transitions in {Lieb}-$5$ {Hubbard} model}, 
      author={Sougata Biswas and Achintyaa and Paramita Dutta},
      eprint={2601.14200},
      archivePrefix={arXiv},
      url={https://arxiv.org/abs/2601.14200}, 
}

@article{ferrari2024altermagnetism,
  title = {Altermagnetism on the {Shastry}-{Sutherland} lattice},
  author = {Ferrari, Francesco and Valent\'{\i}, Roser},
  journal = {Phys. Rev. B},
  volume = {110},
  issue = {20},
  pages = {205140},
  numpages = {11},
  year = {2024},
  month = {Nov},
  publisher = {American Physical Society},
  doi = {10.1103/PhysRevB.110.205140},
  url = {https://link.aps.org/doi/10.1103/PhysRevB.110.205140}
}

@article{sato2024altermagnetic,
  title = {Altermagnetic Anomalous Hall Effect Emerging from Electronic Correlations},
  author = {Sato, Toshihiro and Haddad, Sonia and Fulga, Ion Cosma and Assaad, Fakher F. and van den Brink, Jeroen},
  journal = {Phys. Rev. Lett.},
  volume = {133},
  issue = {8},
  pages = {086503},
  numpages = {6},
  year = {2024},
  month = {Aug},
  publisher = {American Physical Society},
  doi = {10.1103/PhysRevLett.133.086503},
  url = {https://link.aps.org/doi/10.1103/PhysRevLett.133.086503}
}

@article{reimers2024direct,
  title={Direct observation of altermagnetic band splitting in {CrSb} thin films},
  author={Reimers, Sonka and Odenbreit, Lukas and {\v{S}}mejkal, Libor and Strocov, Vladimir N and Constantinou, Procopios and Hellenes, Anna B and Jaeschke Ubiergo, Rodrigo and Campos, Warlley H and Bharadwaj, Venkata K and Chakraborty, Atasi },
  journal={Nat. Commun.},
  volume={15},
  number={1},
  pages={2116},
  year={2024},
  publisher={Nature Publishing Group UK London},
  url={https://doi.org/10.1038/s41467-024-46476-5}
}

@article{li2025topological,
  title={Topological weyl altermagnetism in {CrSb}},
  author={Li, Cong and Hu, Mengli and Li, Zhilin and Wang, Yang and Chen, Wanyu and Thiagarajan, Balasubramanian and Leandersson, Mats and Polley, Craig and Kim, Timur and Liu, Hui},
  journal={Commun. Phys.},
  volume={8},
  number={1},
  pages={311},
  year={2025},
  publisher={Nature Publishing Group UK London},
  url={https://doi.org/10.1038/s42005-025-02232-9}
}

@article{guo2024direct,
  title={Direct and inverse spin splitting effects in altermagnetic {RuO$_2$}},
  author={Guo, Yaqin and others},
  journal={Adv. Sci.},
  volume={11},
  number={25},
  pages={2400967},
  year={2024},
  url = {https://advanced.onlinelibrary.wiley.com/doi/abs/10.1002/advs.202400967},
  publisher={Wiley Online Library}
}

@article{bhowal2024ferroically,
  title = {Ferroically Ordered Magnetic Octupoles in $d$-Wave Altermagnets},
  author = {Bhowal, Sayantika and Spaldin, Nicola A.},
  journal = {Phys. Rev. X},
  volume = {14},
  issue = {1},
  pages = {011019},
  numpages = {19},
  year = {2024},
  month = {Feb},
  publisher = {American Physical Society},
  doi = {10.1103/PhysRevX.14.011019},
  url = {https://link.aps.org/doi/10.1103/PhysRevX.14.011019}
}

@article{lee2024broken,
  title = {Broken Kramers Degeneracy in Altermagnetic {MnTe}},
  author = {Lee, Suyoung and Lee, Sangjae and Jung, Saegyeol and Jung, Jiwon and Kim, Donghan and Lee, Yeonjae and Seok, Byeongjun and Kim, Jaeyoung and Park, Byeong Gyu and \ifmmode \check{S}\else \v{S}\fi{}mejkal, Libor and Kang, Chang-Jong and Kim, Changyoung},
  journal = {Phys. Rev. Lett.},
  volume = {132},
  issue = {3},
  pages = {036702},
  numpages = {7},
  year = {2024},
  month = {Jan},
  publisher = {American Physical Society},
  doi = {10.1103/PhysRevLett.132.036702},
  url = {https://link.aps.org/doi/10.1103/PhysRevLett.132.036702}
}

@article{osumi2024observation,
  title = {Observation of a giant band splitting in altermagnetic {MnTe}},
  author = {Osumi, T. and Souma, S. and Aoyama, T. and Yamauchi, K. and Honma, A. and Nakayama, K. and Takahashi, T. and Ohgushi, K. and Sato, T.},
  journal = {Phys. Rev. B},
  volume = {109},
  issue = {11},
  pages = {115102},
  numpages = {8},
  year = {2024},
  month = {Mar},
  publisher = {American Physical Society},
  doi = {10.1103/PhysRevB.109.115102},
  url = {https://link.aps.org/doi/10.1103/PhysRevB.109.115102}
}

@article{krempasky2024altermagnetic,
  title={Altermagnetic lifting of Kramers spin degeneracy},
  author={Krempask{\`y}, Juraj and {\v{S}}mejkal, L and D’souza, SW and Hajlaoui, M and Springholz, G and Uhl{\'\i}{\v{r}}ov{\'a}, K and Alarab, F and Constantinou, PC and Strocov, V and Usanov, D },
  journal= {Nature},
  volume={626},
  number={7999},
  pages={517--522},
  year={2024},
  url={https://doi.org/10.1038/s41586-023-06907-7},
  publisher={Nature Publishing Group UK London}
}

@article{nag2024GdAlSi,
  title = {{GdAlSi}: An antiferromagnetic topological Weyl semimetal with nonrelativistic spin splitting},
  author = {Nag, Jadupati and others},
  journal = {Phys. Rev. B},
  volume = {110},
  issue = {22},
  pages = {224436},
  numpages = {13},
  year = {2024},
  month = {Dec},
  publisher = {American Physical Society},
  doi = {10.1103/PhysRevB.110.224436},
  url = {https://link.aps.org/doi/10.1103/PhysRevB.110.224436}
}

@article{guo2023spin,
title = {Spin-split collinear antiferromagnets: A large-scale ab-initio study},
journal = {Mater. Today Phys.},
volume = {32},
pages = {100991},
year = {2023},
issn = {2542-5293},
doi = {https://doi.org/10.1016/j.mtphys.2023.100991},
url = {https://www.sciencedirect.com/science/article/pii/S2542529323000275},
author = {Yaqian Guo and Hui Liu and Oleg Janson and Ion Cosma Fulga and Jeroen {van den Brink} and Jorge I. Facio},
}

@article{beida2025chiral,
  title={Chiral split magnons in metallic g-wave altermagnets: insights from many-body perturbation theory},
  author={Beida, Wejdan and {\c{S}}a{\c{s}}{\i}o{\u{g}}lu, Ersoy and Friedrich, Christoph and Bihlmayer, Gustav and Mokrousov, Yuriy and Bl{\"u}gel, Stefan},
  journal={npj Quantum Materials},
  volume={10},
  number={1},
  pages={97},
  year={2025},
  publisher={Nature Publishing Group UK London},
  DOI={10.1038/s41535-025-00818-8},
  url={https://doi.org/10.1038/s41535-025-00818-8}
}

@article{fouet2003planar,
  title = {Planar pyrochlore: A valence-bond crystal},
  author = {Fouet, J.-B. and Mambrini, M. and Sindzingre, P. and Lhuillier, C.},
  journal = {Phys. Rev. B},
  volume = {67},
  issue = {5},
  pages = {054411},
  numpages = {8},
  year = {2003},
  month = {Feb},
  publisher = {American Physical Society},
  doi = {10.1103/PhysRevB.67.054411},
  url = {https://link.aps.org/doi/10.1103/PhysRevB.67.054411}
}

@article{gonzalez2021efficient,
  title = {Efficient Electrical Spin Splitter Based on Nonrelativistic Collinear Antiferromagnetism},
  author = {Gonz\'alez-Hern\'andez, Rafael and \ifmmode \check{S}\else \v{S}\fi{}mejkal, Libor and V\'yborn\'y, Karel and Yahagi, Yuta and Sinova, Jairo and Jungwirth, Tom\'a\ifmmode \check{s}\else \v{s}\fi{} and \ifmmode \check{Z}\else \v{Z}\fi{}elezn\'y, Jakub},
  journal = {Phys. Rev. Lett.},
  volume = {126},
  issue = {12},
  pages = {127701},
  numpages = {6},
  year = {2021},
  month = {Mar},
  publisher = {American Physical Society},
  doi = {10.1103/PhysRevLett.126.127701},
  url = {https://link.aps.org/doi/10.1103/PhysRevLett.126.127701}}

\end{document}